\PassOptionsToPackage{usenames,dvipsnames}{color}
\documentclass[onecolumn,aps,pre,nofootinbib,floatfix,preprintnumbers,amsmath,amssymb,superscriptaddress,10pt,colorlinks,notitlepage]{revtex4-1}
\usepackage{soul}
\usepackage{graphicx}
\usepackage[dvips]{epsfig}
\usepackage{amssymb}
\usepackage{bm}
\usepackage{color}

\usepackage{textcomp}
\usepackage[colorlinks,urlcolor=Black,citecolor=Black,linkcolor=Black]{hyperref}

\usepackage{color}
\usepackage{xcolor}
\usepackage{colortbl}

\usepackage{amsmath}
\usepackage{enumerate}

\usepackage{here}

\newcommand{\blue}[1]{\textcolor{NavyBlue}{#1}}
\definecolor{hellgrau}{rgb}{0.95,0.95,0.95}
\renewcommand{\vec}[1]{\mathbf{{#1}}}
\newcommand{\mean}[1]{\left < #1 \right >}

\newcommand{\be}{\begin{eqnarray}}
\newcommand{\ee}{\end{eqnarray}}
\newcommand{\bse}{\begin{subequations}}
\newcommand{\ese}{\end{subequations}}

\newcommand{\bnum}{\begin{enumerate}}
\newcommand{\enum}{\end{enumerate}}

\newcommand{\bit}{\begin{itemize}}
\newcommand{\eit}{\end{itemize}}

\newcommand{\bc}{\begin{cases}}
\newcommand{\ec}{\end{cases}}

\newcommand{\bpm}{\begin{pmatrix}}
\newcommand{\epm}{\end{pmatrix}}

\newcommand{\bvm}{\begin{vmatrix}}
\newcommand{\evm}{\end{vmatrix}}

\newcommand{\bs}{\boldsymbol}

\newcommand{\p}{\partial}

\newcommand{\abs}[1]{\left | #1 \right |}

\newcommand{\bP}{{\bf P}}

\newcommand{\bu}{{\bf u}}

\makeatletter
\renewcommand\frontmatter@abstractwidth{\dimexpr\textwidth-0.6in\relax}
\makeatother

\begin{document}

% Page header
% \markboth{M.~B\"ar et al.}{Self-Propelled Rods}

% Title
%\title{Self-Propelled Rods as a Paradigm of Active Matter}
\title{Self-Propelled Rods:~Insights and Perspectives for Active Matter}

%Authors, affiliations address in journal format
\author{Markus B\"ar}
\affiliation{Physikalisch-Technische Bundesanstalt, Berlin, Germany, D-10587}

\author{Robert Gro{\ss}mann}
\affiliation{Institute for Physics and Astronomy, University of Potsdam, Potsdam, Germany, D-14476}

\author{Sebastian Heidenreich}
\affiliation{Physikalisch-Technische Bundesanstalt, Berlin, Germany, D-10587}

\author{Fernando Peruani}
\affiliation{Laboratoire J.A. Dieudonn\'{e}, Universit\'{e} C\^{o}te d'Azur, Nice, France, F-06108}

\begin{abstract}

A wide range of experimental systems including gliding, swarming and swimming bacteria, in-vitro motility assays as well as shaken granular media are commonly described as self-propelled rods. 
Large ensembles of those entities display a large variety of self-organized, collective phenomena, including formation of moving polar clusters, polar and nematic dynamic bands, mobility-induced phase separation, topological defects and mesoscale turbulence, among others. 
Here, we give a brief survey of experimental observations and review the theoretical description of self-propelled rods. 
Our focus is on the emergent pattern formation of ensembles of \textit{dry} self-propelled rods governed by short-ranged, contact mediated interactions and their \textit{wet} counterparts that are also subject to long-ranged hydrodynamic flows. 
Altogether, self-propelled rods provide an overarching theme covering many aspects of active matter containing well-explored limiting cases. 
Their collective behavior not only bridges the well-studied regimes of polar self-propelled particles and active nematics, and includes active phase separation, but also reveals a rich variety of new patterns.  

\end{abstract}

% % keywords, separated by comma, no full stop, lowercase
% \begin{keywords}
% active matter, self-propelled rods, collective motion, stochastic pattern formation far from thermodynamic equilibrium 
% \end{keywords}

% create the title
\maketitle

%%%%%%%%%%%%%%%%%%%%%%%%%%%%%%%%%%%%%%%%%%%%%%%%%%%%%%%
%%%%%%%%%%%%%%%%%%%%%%%%%%%%%%%%%%%%%%%%%%%%%%%%%%%%%%%

\setcounter{page}{1}

%%%%%%%%%%%%%%%%%%%%%%%%%%%%%%%%%%%%%%%%%%%%%%%%%%%%%%%
%%%%%%%%%%%%%%%%%%%%%%%%%%%%%%%%%%%%%%%%%%%%%%%%%%%%%%%

\section{INTRODUCTION}
\label{sec:intro}

% the topic of this review
%
The topic of self-propelled rods grew out of the non-equilibrium extension of a well-known showcase of liquid crystal physics~--~the study of the equilibrium phases of diffusive passive rods with steric repulsion interactions~--~by assuming that such rods exhibit persistent self-propelled motion.
Nowadays, self-propelled rods have become one prominent example of so-called active matter along with related topics like active nematics or collective motion of polar active particles~\cite{vicsek_collective_2012,marchetti_hydrodynamics_2013}.
As a result, many experimental and theoretical studies have been carried out in recent years.
In this review, we will give a brief account on progress in the understanding of self-propelled rod systems and emphasize common aspects connecting many diverse applications.

% passive liquid crystals
% 
The equilibrium physics of liquid crystals is a cornerstone of condensed matter theory~\cite{chaikin_principles_2000}. 
The structures and phases formed by these mesomorphic materials~--~combining properties of fluids and solids~--~are essentially classified on the basis of their symmetries and conservation laws~\cite{deGennes_physics_1993}. 
Liquid crystals are composed of anisotropic entities such as elongated, rod-like molecules~\cite{doi_theory_1986}; in simplified terms, one can think of them as \textit{passive rods}. 
Phase transitions towards nematic and smectic order associated with critical phenomena as well as nontrivial dynamics of topological defects, controlled externally by temperature, particle density, solvent dynamics via external shear or electromagnetic fields, for example, were studied extensively from the perspective of statistical mechanics, nonlinear dynamics and pattern formation in the past~\cite{deGennes_physics_1993}. 
This knowledge is now used in technological applications like liquid crystal displays.

% touch upon self-propelled particles 
%
The phase transition from a disordered collection of passive rods interacting via volume exclusion towards uniaxial orientational ordering as a function of the rod density was first understood by Onsager theoretically within a simplified model of hard rods~\cite{onsager_effects_1949,kayser_bifurcation_1978}:~as two, close-by rods align in a parallel or anti-parallel fashion, i.e.~\textit{uniaxial}, \textit{apolar} or \textit{nematic} alignment, the excluded volume is minimized thereby maximizing positional entropy. 
Accordingly, nematic orientational order emerges as the particle density is increased. 
How will this ordering transition change if individuals possess an internal energy reservoir that enables them to move persistently along their body axis in a unidirectional fashion~--~thereby breaking the apolar symmetry of the interactions in the corresponding equilibrium system~--~by means of a self-propulsion mechanism, i.e.~if they are \textit{active particles} or more precisely \textit{self-propelled rods}~(see box below for a definition)?

% active matter and active particles in general
%
In general, a particle is called \textit{active} or \textit{self-propelled} if it can transduce free energy into persistent motion~\cite{ramaswamy_mechanics_2010}. 
Large ensembles of these active particles have been labeled as active matter~\cite{ramaswamy_mechanics_2010,vicsek_collective_2012,marchetti_hydrodynamics_2013}. 
Due to nonequilibrium nature of self-propulsion at the particle level, all forms of active matter exhibit a large variety of collective behaviors and pattern formation processes that are fundamentally different from their equilibrium counterparts as these system evade the constraints of energy conservation, the action-reaction principle or standard fluctuation-dissipation theorems. 
Examples of active matter include all forms of living matter ranging from the cellular self-organization via dynamics of biological tissues and bacterial aggregation to fish schools, bird flocks and herds of large mammals~\cite{vicsek_collective_2012,marchetti_hydrodynamics_2013}. 
First attempts to engineer active matter systems artificially include diverse fields from active colloids~\cite{zottl_emergent_2016} to granular active matter~\cite{yamada_coherent_2003,aranson_patterns_2006,narayan_long_2007}. 
As a result, the field of active matter has experienced a rapid growth throughout many disciplines:~statistical physics and soft matter theory as well as nonlinear dynamics and fluid dynamics have contributed to the development and analysis of model systems as well as to the general understanding of emergent phenomena in active materials. 
However, an encompassing active matter theory framework, analogous to thermodynamics for equilibrium systems, has not been established yet.

\vspace{0.35cm}
\fboxsep0pt
\noindent
\colorbox{hellgrau}{
\begin{minipage}{0.988\textwidth}
\vspace{0.4cm}

\begin{center}
		
\parbox[c]{0.9\textwidth}{

	\vspace{0.16cm}

	\normalsize{\bfseries{\sffamily{\blue{\bfseries{\hspace*{-0.02cm}SELF-PROPELLED RODS}}}}}

	\normalsize
	\normalfont

	\vspace{0.3cm}

	Self-propelled rods are a class of active particles:~they are persistently self-propelled in a unidirectional~(\textit{polar}) way along their body axis. 
	On the basis of their interaction, two subclasses are distinguished from each other. 

	\vspace{0.2cm}

	\blue{\sffamily{\bfseries{dry self-propelled rods. }}}
	Self-propelled rods are spatially extended objects with a well-defined aspect ratio~$\chi$~(length-to-width ratio) persistently moving in a dissipative, fluctuating environment interacting by short-ranged, repulsive interactions dictated by their shape. 
	The interaction is, at the microscopic level, \textit{apolar} in the sense that it is invariant with respect to inversions of the body axis of a rod~(uniaxial symmetry). 
	Limiting cases include self-propelled discs~($\chi = 1$) that may display so-called motility-induced phase separation~\cite{cates_mips_2015} as well as dry active nematics for vanishing persistent self-propulsion displaying isotropic-nematic ordering locally and giant number fluctuations~\cite{chate_simple_2006,mishra_active_2006}; for reviews, see~\cite{ramaswamy_mechanics_2010,marchetti_hydrodynamics_2013,cates_mips_2015}. 
	Away from these limits, self-propelled rods display a larger variety of collective behaviors, including the formation of moving polar clusters, giant aggregates, polar bands and laning. 

	\vspace{0.2cm}

	\blue{\sffamily{\bfseries{wet self-propelled rods. }}}
	In addition to their characteristic short-ranged repulsion, wet self-propelled rods in fluid environments are coupled via the surrounding flow field of the solvent. 
	This long-ranged hydrodynamic interaction together with the incompressibility of the flow changes fundamentally the physics of the problem, leading to the emergence of irregular vortex dynamics and topological defects, which are characteristic of mesoscale turbulence~\cite{wensink_meso-scale_2012,doostmohammadi_active_2018}. 

}

\end{center}

\vspace{0.1cm}
\end{minipage}
}
\vspace{0.2cm}

Interacting self-propelled rods are a paradigm of active matter as a broad variety of active systems belongs to this class. 
The most frequent realization of self-propelled rods in nature are bacteria gliding on surfaces or swimming in fluids, cf.~a recent review article on the statistical physics view of bacterial swarming~\cite{beer_statistical_2019}. 
The typical rod-like shape of bacteria is maintained by a feedback between cell curvature and cytoskeletal localization~\cite{ursell_rod_2014}. 
Inside single cells, biopolymers such as actin or microtubules are often self-propelled by molecular motors such as myosin, kinesin or dynein. 
In-vitro, motility assays of biopolymers propelled by motors attached to a surface allow well-defined studies of actin filaments~\cite{schaller_polar_2010,huber_emergence_2018} and microtubuli~\cite{sumino_large-scale_2012} behaving like self-propelled rods. 
Artificial particles behaving as self-propelled rods are shaken cylindrical granular particles with unequal weight distribution~\cite{kudrolli_swarming_2008,kumar_flocking_2014} or chemically driven rod-shaped Janus particles~\cite{paxton_catalytic_2004}.

\begin{figure}[t]
\begin{center}
\includegraphics[width=\textwidth]{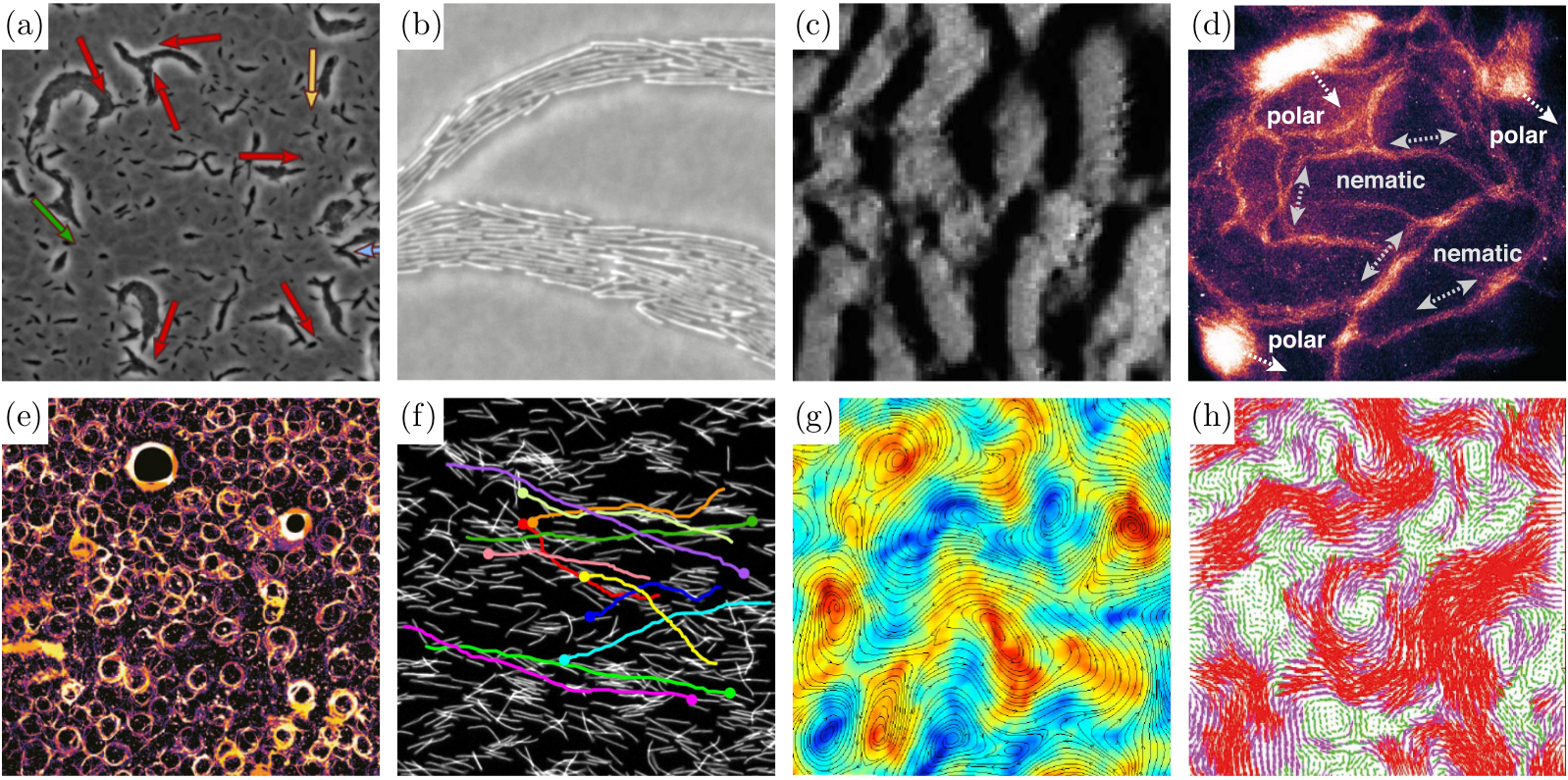}	
\end{center}
\vspace{-0.3cm}
\caption{Experimental examples of collective dynamics and pattern formation in self-propelled rods: (a)~polar clusters in sub-monolayer of \textit{Myxococcus xanthus}~(from Peruani \textit{et al.}~\cite{peruani_collective_2012}); (b)~nematic bands in \textit{Paenibacillus dendritiformis}~(from Be{\textquotesingle}er~\textit{et al}.~\cite{beer_periodic_2013}); (c)~standing waves in multilayers of \textit{Myxococcus xanthus}~(from B\"orner \textit{et al.}~\cite{boerner_rippling_2002}); (d)~coexisting polar clusters and nematic bands in an actin-myosin motility assay~(from Huber \textit{et al.}~\cite{huber_emergence_2018}); (e)~vortex arrays in a tubulin-dynein motility assay~(from Sumino \textit{et al.}~\cite{sumino_large-scale_2012}); (f)~long-range nematic order in filamentous \textit{Escherichia coli}~(from Nishiguchi~\textit{et~al.}~\cite{nishiguchi_long_2017}); (g)~mesoscale turbulence in densely packed \textit{Bacillus subtilis} in a solvent fluid~(from Dunkel \textit{et al.}~\cite{dunkel_fluid_2013}); (h)~mesoscale turbulence in \textit{Serratia marcescens}~(from Rabani \textit{et al.}~\cite{rabani_collective_2013}). }
\label{fig:phenomena}
\end{figure}

A large variety of collective phenomena has been reported in the experimental realizations of self-propelled rods mentioned above. 
The formation of polar clusters with a characteristic size distribution within which rods move in parallel and related giant number fluctuations were found in experiments employing monolayers of swarming \textit{Bacillus subtilis}~\cite{zhang_agent_2017} and in \textit{Myxococcus xanthus} mutants~\cite{peruani_collective_2012,thutupalli_directional_2015,starruss_pattern_2012}, see Fig.~\ref{fig:phenomena}a.
The formation of narrow, elongated, high-density regions called bands along which rod-shaped bacteria move was observed in the wild type of \textit{Myxococcus xanthus}~\cite{starruss_pattern_2012,thutupalli_directional_2015} and in \textit{Paenibacillus dendritiformis}~\cite{beer_periodic_2013}, see Fig.~\ref{fig:phenomena}b.
Multilayers of nematically ordered myxobacteria give rise to counter-propagating density waves known as rippling that has been linked to cell reversals due to chemical signaling~\cite{boerner_rippling_2002,igoshin_pattern_2001,harvey_continuum_2013}, see Fig.~\ref{fig:phenomena}c.
Studies in motility assays have revealed traveling polar bands~\cite{schaller_polar_2010} as well as moving polar cluster and stationary nematic bands~\cite{huber_emergence_2018} for the case of actin and myosin, see Fig.~\ref{fig:phenomena}d, as well as vortex arrays for the tubulin-dynein system~\cite{sumino_large-scale_2012}, see Fig.~\ref{fig:phenomena}e. 
Long-range nematic order was reported in experiments with fibroblasts \cite{li_on_2017}, on monolayers of spindle-shaped cells \cite{duclos_perfect_2014} and for filamentous swimming \textit{Escherichia coli}~\cite{nishiguchi_long_2017}, see Fig.~\ref{fig:phenomena}f.
Mesoscale turbulence governed by irregular vortex dynamics was found to be the most typical pattern for many suspensions of swimming bacteria, e.g.~\textit{Bacillus subtilis} \cite{sokolov_concentration_2007,sokolov_reduction_2009,wensink_meso-scale_2012,dunkel_fluid_2013,benisty_antibiotic_2015,ilkanaiv_effect_2017}, see Fig.~\ref{fig:phenomena}g, \textit{Paenibacillus dendritiformis}~\cite{zhang_swarming_2009} and \textit{Serratia marcescens}~\cite{rabani_collective_2013,ariel_collective_2018}, see Fig.~\ref{fig:phenomena}h.

Several of the collective patterns found in experimental self-propelled rod systems were first predicted by early models, namely the formation of polar clusters~\cite{peruani_nonequilibrium_2006}, the emergence of (nematic) bands~\cite{ginelli_large_2010} as well as the possible existence of long-ranged nematic order in two spatial dimensions~\cite{peruani_mean_2008,ginelli_large_2010}. 
Those early studies revealed that an essential aspect of the microscopic dynamics that has a strong impact on the emergent macroscopic patterns  is whether rods can crawl over each other. 
If this occurs~--~a situation commonly observed in experiments with swarming bacteria in quasi-two-dimensional systems~--~the dynamics of the system can be reduced to that of self-propelled point-like particles with nematic interactions\footnote{We refer the reader to \cite{vicsek_novel_1995} for the first systematic investigation of self-propelled point particles with polar alignment from a physical point of view. } \cite{peruani_mean_2008,ginelli_large_2010}.
Then, the formation of dynamic nematic bands and long-ranged nematic order (currently subject of debate) are observed. 
On the other hand~--~if rods cannot cross each other~--~the physics of the problem is fundamentally different. 
Polar moving clusters~\cite{peruani_nonequilibrium_2006} and even highly dynamic, polar bands \cite{weitz_selfpropelled_2015,abkenar_collective_2013} emerge and dominate the large-scale dynamics of the system.  
There are other microscopic details that affect the large-scale properties of these nonequilibrium active systems. 
From the many models developed to understand the variety of collective patterns observed in motility assays experiments~\cite{sumino_large-scale_2012,huber_emergence_2018,kraikivski_enhanced_2006,weber_random_2015,suzuki_polar_2015}, it is evident that the flexibility of the rods~~\cite{sumino_large-scale_2012,huber_emergence_2018} can affect the large-scale properties of these systems.  
Finally, models~\cite{dunkel_fluid_2013,grossmann_vortex_2014,heidenreich_hydrodynamic_2016} developed to shed light on bacterial mesoscale turbulence~\cite{wensink_meso-scale_2012,doostmohammadi_active_2018} indicate that hydrodynamic interactions also have a strong impact on the macroscopic, emergent dynamics of these systems.

Beyond the relevance of self-propelled rod systems to understand a large variety of experimental systems, these systems provide an overarching theme in the broad context of active matter, containing most classes and collective phenomena reported in active systems~\cite{ramaswamy_mechanics_2010,vicsek_collective_2012,marchetti_hydrodynamics_2013,cates_mips_2015}. 
In the limit of vanishing active propulsion and for mobility coefficients and fluctuations obeying the standard fluctuation-dissipation relation, we recover Onsager theory for passive rods~\cite{onsager_effects_1949,kayser_bifurcation_1978}. 
By restoring active propulsion and in the limit of aspect ratio one, corresponding to isotropic interactions, rods become discs and motility induced phase separation~(MIPS)~\cite{cates_mips_2015} is expected to be observed.
As the aspect ratio is increased, anisotropic interactions among the rods lead to active torques and alignment, in turn destabilizing MIPS~\cite{shi_self_2018,grossmann_particle_2019}.  
The presence of anisotropic  interactions, as mentioned above, can lead, counterintuitively, to the emergence of both, nematic  and polar  patterns \cite{weitz_selfpropelled_2015,abkenar_collective_2013,shi_self_2018,grossmann_particle_2019}. 
Thus, the symmetry of the emergent order is a dynamical property of the system in self-propelled rods and not a consequence of the~(built-in) interaction potential symmetry like in Vicsek-type models~\cite{vicsek_novel_1995,ginelli_large_2010}.  
Finally, in the limit of vanishing active propulsion and for non-thermal fluctuations~[cf.~\cite{chate_simple_2006,mishra_active_2006,peshkov_nonlinear_2012,grossmann_mesoscale_2016}], the dynamics of self-propelled rods reduces to that of active nematics~\cite{ramaswamy_mechanics_2010,marchetti_hydrodynamics_2013,doostmohammadi_active_2018}.  
It is important to stress that self-propelled rods exhibit a much richer phenomenology than active nematic systems~--~a mapping onto active nematics is possible in limiting cases only. 
The observation of polar patterns in self-propelled rod systems or MIPS provides a good hint about how profound the differences between these classes of systems are.  
In terms of hydrodynamic equations, the difference between self-propelled rods and active nematics can be understood as follows:~the large-scale properties of active nematics are described by two slowly varying fields, namely the local density and nematic order parameter~\cite{marchetti_hydrodynamics_2013,baskaran_self_2012}, however, a description of the large-scale properties of self-propelled rods requires, in general, the use of three fields~--~local density, nematic order as well as polar order parameter~--~and it involves convective mass transport.

This review is organized as follows. 
Section~\ref{sec:drySPR} deals with dry self-propelled rods, i.e.~rods that exhibit short range interactions. 
Results on self-propelled rods with steric repulsion as well as on self-propelled rods with nematic alignment interaction are discussed in detail. 
In Section~\ref{sec:wetSPR}, we consider self-propelled rods embedded in a fluid and discuss the cases of short range polar and nematic alignment in combination with long-range hydrodynamic coupling. 
In Sections~\ref{sec:drySPR} and~\ref{sec:wetSPR} kinetic theories and derivations of continuum hydrodynamic equations from the stochastic Langevin-type dynamics of single self-propelled rods, respectively single rod-shaped swimmers in a fluid, are provided. 
The paper closes with a summary and a brief sketch of future research topics.

%%%%%%%%%%%%%%%%%%%%%%%%%%%%%%%%%%%%%%%%%%%%%%%%%%%%%%%
%%%%%%%%%%%%%%%%%%%%%%%%%%%%%%%%%%%%%%%%%%%%%%%%%%%%%%%

\section{DRY SELF-PROPELLED RODS}
\label{sec:drySPR}

\subsection{Nonequilibrium dynamics of dry self-propelled rods}

The term~\textit{self-propelled rod} was introduced in physics~in 2006~\cite{peruani_nonequilibrium_2006} for rod-shaped objects with anisotropic repulsive interactions that drive their motion actively through a dissipative medium. 
As self-propulsion requires a perpetual energy influx at the particle level, the dynamics is far from thermodynamic equilibrium, thereby allowing for a plethora of nontrivial self-organization and pattern formation phenomena that could not emerge in ensembles of passive rods.

\begin{figure}[t]
\begin{center}
\vspace{0.2cm}
\includegraphics[width=0.72\textwidth]{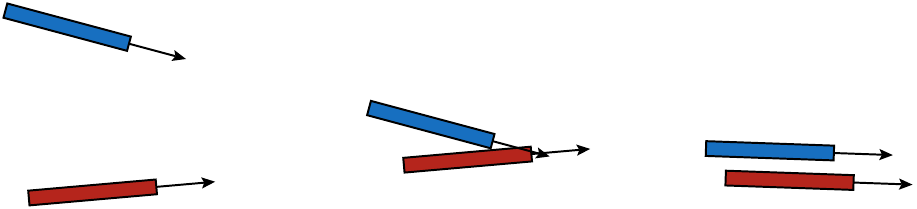}
\end{center}
\vspace{-0.3cm}
\caption{A binary collision of self-propelled rods:~the interplay of active motion and anisotropic repulsion effectively implies alignment of their velocities. }
\label{fig:nem_al_illust}
\end{figure}

To get an intuition for this novel physics, consider a binary collision as shown in Fig.~\ref{fig:nem_al_illust}:~rods colliding under an acute angle align their direction of motion parallel due to the combined effect of anisotropic repulsion and self-propulsion.
Once aligned, they move together in the same direction for some time.
This simple picture illustrates the interrelation of self-propulsion, anisotropic interactions, density instabilities, the emergence of local order as well as the tendency towards clustering.

The binary collision reveals yet another important nonequilibrium aspect of the dynamics:~as rods move at a non-vanishing, characteristic speed before and after the collision but aligned their direction of motion, the dynamics does not conserve momentum. 
The momentum difference is transferred to the environment, e.g.~a surrounding fluid or a surface that acts effectively as a momentum sink.

We classify self-propelled rod systems where this type of contact-mediated, anisotropic repulsive interaction is dominant as \textit{dry self-propelled rods} which we review in this section.
Their wet counterparts~--~ensembles of self-propelled rods with long-ranged hydrodynamic interactions mediated by a solvent flow~--~are the topic of Section~\ref{sec:wetSPR}.

\subsubsection{Equations of motion}
\label{subs:lang_dyn}

The full nonequilibrium dynamics of an ensemble of~$N$ dry self-propelled rods in two spatial dimensions with positions~$\vec{r}_j(t)$ and orientations~$\varphi_j(t)$ is described by the following set of Langevin equations: 
\begin{subequations}
\label{eqn:mot}
\begin{align}
	\dot{\vec{r}}_j(t) &= \hat{\boldsymbol{\mu}} \! \left[ \varphi_j \right] \!\!\: \cdot \!\!\: \left [ \vec{F}^{\mbox{\tiny{(act)}}}_j \! + \sum_{k \neq j} \, \vec{f}_2 \! \left( \vec{r}_k - \vec{r}_j; \varphi_k, \varphi_j \right) \right ] \! + \!\, \boldsymbol{\eta}_j(t), \\ 
	\dot{\varphi}_j(t) &= \mu_{\varphi} \sum_{k \neq j}\, m_2 \! \left( \vec{r}_k - \vec{r}_j; \varphi_k, \varphi_j \right) +\!\!\: \sqrt{2 D_{\varphi} } \!\; \xi_j \! \left( t \right) \! .
\end{align}
\end{subequations}
Central elements determining the dynamics are: 
\begin{itemize}
	\item[--] \textit{anisotropic repulsion}:~Self-propelled rods repel each other, as reflected by the anisotropic, binary force~$\vec{f}_2$. 
	\item[--] \textit{aligning torque}:~Self-propelled rods align their body axis in an uniaxial way upon collision~(Fig.~\ref{fig:nem_al_illust}) described by the torque~$m_2$ that explicitly depends on rod shape.   
	\item[--] \textit{self-propulsion}:~A typical model~\cite{peruani_nonequilibrium_2006,wensink_aggregation_2008,baskaran_enhanced_2008,shi_self_2018} for the persistent, ballistic motion of rods at short time scales contains a force oriented along the long body axis~$\vec{e}_{\parallel} \! \left[ \varphi_j \right]$ with a nonzero average amplitude:~$\vec{F}^{\mbox{\tiny{(act)}}}_j \!\!\:=\!\!\: F_0 \!\: \vec{e}_{\parallel} \! \left[ \varphi_j \right]$. 
	\item[--] \textit{fluctuations}:~Noises with the amplitudes~$D_{{\parallel}}$, $D_{\perp}$ and $D_{\varphi}$ account for inherently stochastic self-propulsion or spatial heterogeneities of the substrate on which rods move. Consequently, these parameters are \textit{independent} and \textit{differ} from the corresponding diffusion coefficients of passive rods in a fluid. Translational fluctuations are generally anisotropic:
\begin{align}
	\label{eqn:def:noise_corr}
 \boldsymbol{\eta}_j(t) \!\!\: = \!\!\: \sqrt{2D_{\parallel}} \!\: \vec{e}_{\parallel} \!\!\: \! \left [ \varphi_j \right ] \!\!\: \eta_{\parallel,j}(t) \!\!\:+\!\!\: \sqrt{2D_{\perp}} \!\: \vec{e}_{\perp} \!\!\: \! \left [ \varphi_j \right ] \!\!\: \eta_{\perp,j}(t). 
\end{align}
	\item[--] \textit{dissipation}:~The mobilities~$\hat{\boldsymbol{\mu}} \! \left[ \varphi_j \right]$ and~$\mu_{\varphi}$ are interpreted as inverse friction coefficients. They depend on the rods' aspect ratio. A standard fluctuation-dissipation relation does usually not hold and the equipartition theorem does not apply to the nonequilibrium dynamics of self-propelled rods, e.g.~if friction and fluctuations are of nonthermal origin. 
\end{itemize}
A characteristic feature of self-propelled rods is the symmetry of their dynamics:~the interaction force and torque, dissipation as well as fluctuations obey an uniaxial symmetry, i.e.~they are invariant if head and tail of a rod are interchanged~($\varphi_j \rightarrow \varphi_j + \pi$).  
The polar self-propulsion force breaks this symmetry. 
Accordingly, the emergence of polar patterns is inherently related to this nonequilibrium driving at the particle level breaking uniaxial symmetry.

\subsubsection{Self-propelled rods within the realm of dry active matter}

\begin{table}[t]
\caption{Classification of \textit{self-propelled rods} within the realm of \textit{dry active matter}. }
\label{tab:conf_dry2}
\begin{center}
	\begin{tabular}{ c || c | c }
	& \cellcolor{hellgrau} & \cellcolor{hellgrau} \\
 & \cellcolor{hellgrau} \textbf{isotropic shape} & \cellcolor{hellgrau} $\quad$ \textbf{anisotropic shape} $\quad$ \\
        & \cellcolor{hellgrau} & \cellcolor{hellgrau} \\
  \hline 
  \hline
   \cellcolor{hellgrau} & & \\
   \cellcolor{hellgrau} $\,$ \textbf{persistent self-propulsion} $\,$ & $\,$ \parbox{4.15cm}{\centering self-propelled discs ($d=2$); \\ self-propelled spheres ($d=3$)} $\,$ &  \textit{self-propelled rods}  \\
   %\cellcolor{hellgrau} $\,$  $\,$ & $\,$ self-propelled spheres ($d=3$) $\,$ &  \\
   \cellcolor{hellgrau} & & \\
  \hline
   \cellcolor{hellgrau} & & \\
   \cellcolor{hellgrau} \textbf{undirected driving} & granular spheres & $\,$ dry active nematics $\,$ \\
   \cellcolor{hellgrau} & & \\
  \hline 
\end{tabular} 
\end{center}
\end{table}

Self-propelled discs (in two spatial dimensions, $d=2$) and self-propelled spheres (in three dimensions, $d=3$) as well as anisotropic particles without directed self-propulsion are limiting cases of self-propelled rods, in turn representing well-studied sub-classes of dry active matter, summarized in Table~\ref{tab:conf_dry2}.

In the limit of spherical particles, sometimes also called \textit{active Brownian particles}\footnote{The term \textit{active Brownian particles} is sometimes used synonymously in a more general context for self-propelled motion far from equilibrium~\cite{romanczuk_active_2012}. }, the aligning torque vanishes assuming that interfacial friction is negligible. 
Consequently, orientational order cannot emerge. 
Ensembles of self-propelled discs may, however, phase-segregate at high levels of activity and exhibit MIPS~\cite{cates_mips_2015}. 
For aspect ratios close to one, the phenomenology displayed by self-propelled rods is related to ensembles of active dumbbells~\cite{cugliandolo_phase_2017}.

If the self-propulsion force tends to zero but particle shape remains anisotropic, the \textit{dry active nematics} class is recovered~\cite{chate_simple_2006,mishra_active_2006,marchetti_hydrodynamics_2013}. 
The stochastic back and forth motion at the particle level implies diffusive dynamics on all timescales in contrast to persistent propulsion of self-propelled rods. 
If stochastic reversals of the self-propulsion force, as observed for the soil bacterium \textit{Myxococcus xanthus}~\cite{munoz_myxo_2016} are included into the model defined by Eq.~\ref{eqn:mot:part}, the dynamics reduces to dry active nematics in the limit of high reversal rates~\cite{grossmann_mesoscale_2016}.

\subsection{Emergent patterns in ensembles of dry self-propelled rods}
\label{sec:dry_pheomnea}

Central parameters which control pattern formation of self-propelled rods are the aspect ratio~\cite{peruani_nonequilibrium_2006}, the overall rod density as well as their their softness, i.e.~the strength of self-propulsion with respect to repulsion~\cite{shi_self_2018}.  
In the following, we first consider cases where repulsion is dominant and turn to the limit of strong self-propulsion afterwards.

\subsubsection{Weak self-propulsion~--~self-propelled rods that push each other}
\label{sec:rod_ve}

\begin{figure}[b]
\begin{center}
\includegraphics[width=0.7\textwidth]{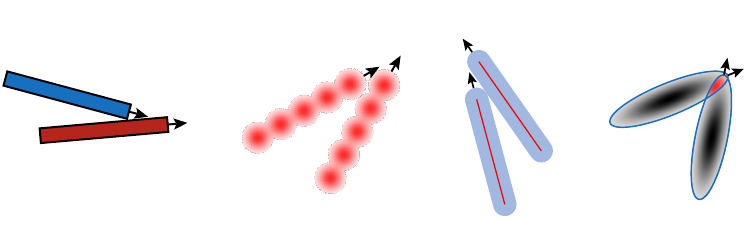}
\end{center}
\vspace{-0.3cm}
\caption{Selection of models for the interaction of self-propelled rods:~representation by rectangles~\cite{peruani_nonequilibrium_2006}, chains of spheres~\cite{wensink_emergent_2012}, spherocylinders~\cite{shi_self_2018} or anisotropic phase fields~\cite{grossmann_particle_2019}. } 
\label{fig:dry_hard_models}
\end{figure}

Self-propelled rods that push each other exert forces and torques to be derived from a specific model that enter into Eq.~\ref{eqn:mot}. 
Some variants are illustrated in Fig.~\ref{fig:dry_hard_models}, e.g.~representing rods by rectangular~\cite{peruani_nonequilibrium_2006} or elliptical objects~\cite{theers_clustering_2018,bott_isotropic_2018,grossmann_particle_2019}, spherocylinders~\cite{shi_self_2018}, needles~\cite{sambelashvili_dynamics_2007} or chains of spheres~\cite{wensink_aggregation_2008,wensink_emergent_2012,abkenar_collective_2013,weitz_selfpropelled_2015}. 
All models have in common that interactions are purely repulsive and short-ranged. 
Some models assume pair-wise repulsion along the axis of the shortest distance~\cite{mccandlish_spontaneous_2012,kuan_hysteresis_2015}, others a potential energy penalizing overlap~\cite{peruani_nonequilibrium_2006,grossmann_particle_2019}.

Usually, there is a trade-off between the accuracy of the description and numerical efficiency:~representing rods as a chain of spheres is a popular and simple model~\cite{wensink_aggregation_2008,wensink_emergent_2012,abkenar_collective_2013,weitz_selfpropelled_2015}, however, this substructure may induce shear friction or blocking of close-by, aligned rods that would not be present if rods could slide along each other during a collision. 
Typically, soft-interactions are numerically simpler to handle than hard repulsion potentials or volume exclusion since forces may become arbitrarily large if the potential energy diverges for small inter-rod distances in the latter case.

\begin{figure}[t]
\begin{center}
\includegraphics[width=\textwidth]{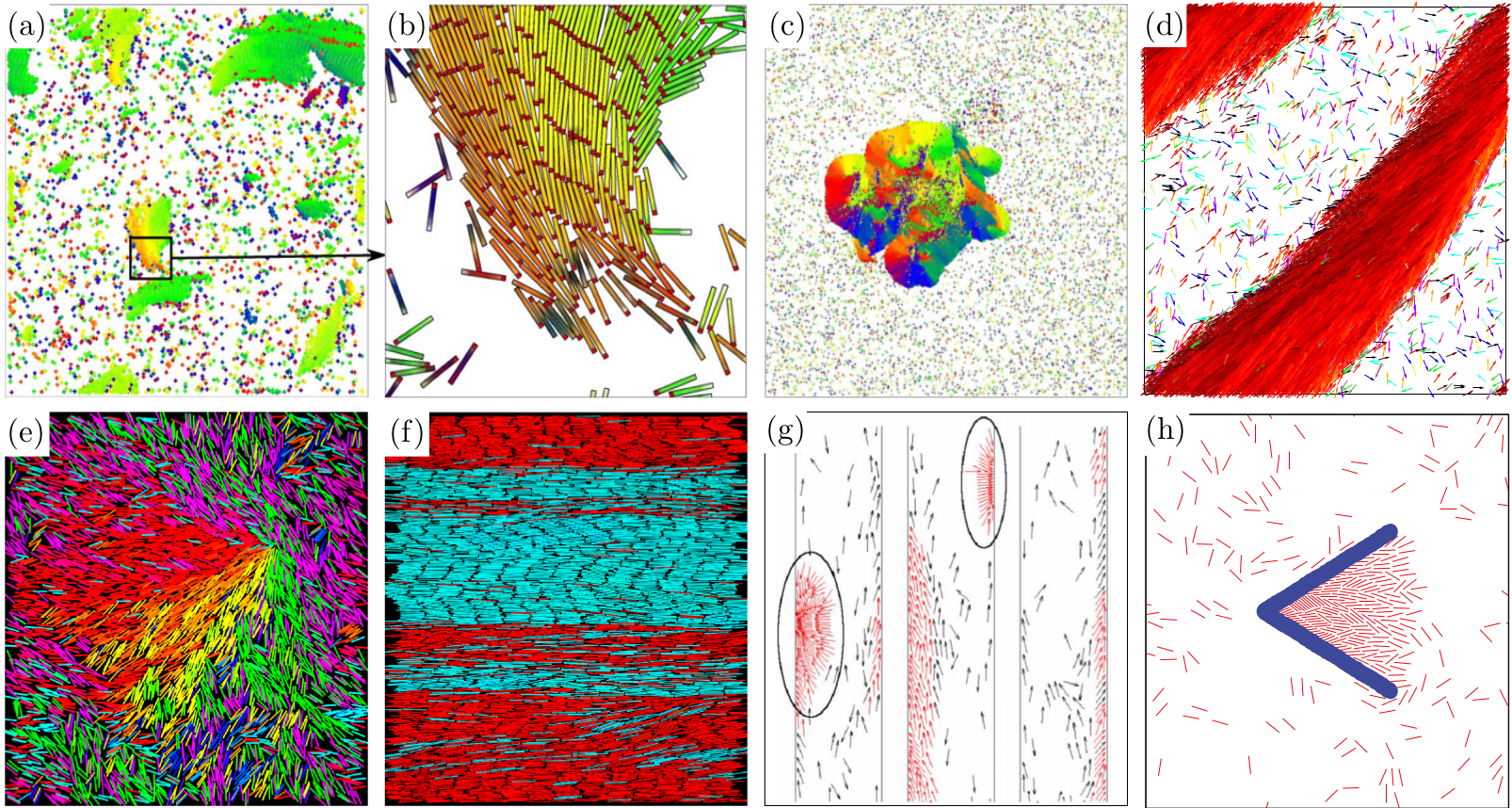}
\end{center}
\vspace{-0.3cm}
\caption{Phenomenology of dry self-propelled rods where the dominant interaction is short-ranged, anisotropic repulsion. (a)~Polar clusters with local smectic order as highlighted by the image section~(b); (c) giant aggregates [(a)-(c) from Weitz~\textit{et~al.}~\cite{weitz_selfpropelled_2015}]; (d) polar bands (from Abkenar \textit{et al.}~\cite{abkenar_collective_2013}); (e) nematic chaos, particularly showing a nematic defect; (f)~laning at high density [(e)-(f) from Shi \textit{et al.}~\cite{shi_self_2018}]; (g) accumulation of self-propelled rods at walls (from Wensink \textit{et al.}~\cite{wensink_aggregation_2008}); (h) a wedge-shaped trap for self-propelled rods (from Kaiser \textit{et al.}~\cite{kaiser_capturing_2013}). }
\label{fig:phenom:dry_theory}
\end{figure}

First numerical works on self-propelled rods~\cite{peruani_nonequilibrium_2006} revealed nonequilibrium clustering above a critical aspect ratio, i.e.~for long rods~[Figs.~\ref{fig:phenom:dry_theory}(a,b)] at a critical density that is, notably, smaller than the percolation threshold and the isotropic-nematic transition expected in two-dimensional equilibrium nematics~\cite{delasheras_two_2013}.
Due to the inherent tendency of self-propelled rods to move in parallel after a collision~(Fig.~\ref{fig:nem_al_illust}), clusters are polar and motile, and may have a smectic~(layered) substructure~[Figs.~\ref{fig:phenom:dry_theory}(b)].  
This type of \textit{aggregation driven by anisotropic repulsion and self-propulsion} is observed in a plethora of microbiological systems~\cite{zhang_collective_2010,peruani_collective_2012,jeckel_learning_2019}.

At the mesoscale, clusters may organize themselves into system-spanning polar bands associated with macroscopic polar order~[Fig.~\ref{fig:phenom:dry_theory}(d)] in simulations with periodic boundary conditions~\cite{abkenar_collective_2013,weitz_selfpropelled_2015,shi_self_2018,grossmann_particle_2019}. 
These structures are not stable as the system size is increased but are found to break apart above a critical system size, and disordered, high density aggregates~[Fig.~\ref{fig:phenom:dry_theory}(c)] form instead within which polar clusters of rods exert nonequilibrium stresses and torques on each other~\cite{weitz_selfpropelled_2015}.  
From time to time, an aggregate looses macroscopic fractions of its mass via the ejection of polar clusters as highlighted in Fig.~\ref{fig:phenom:dry_theory}(c). 
The physics of these aggregates thus differs crucially from the high-density drops which emerge during MIPS in self-propelled discs~\cite{cates_mips_2015} that are due to a density-dependent slow-down in high-density areas. 
This type of quorum sensing interaction was explicitly addressed in the context of self-propelled rods in~\cite{velasco_collective_2018}.

At high densities, a plethora of collective states were found, including~(percolating) turbulence~\cite{wensink_emergent_2012,shi_self_2018} and nematic chaos~\cite{shi_self_2018}~--~highly dynamic regimes where defects move and reorganize constantly~[Fig.~\ref{fig:phenom:dry_theory}(e)]~--~as well as laning~\cite{wensink_emergent_2012,mccandlish_spontaneous_2012,abkenar_collective_2013,kuan_hysteresis_2015} represented in Fig.~\ref{fig:phenom:dry_theory}(f), i.e.~states which display nematic order at the global scale but local polar order, active smectics~\cite{romanczuk_smectic_2016} and jammed states together with reentrant fluidization~\cite{kuan_hysteresis_2015}. 
As many of these patterns are genuinely novel nonequilibrium phenomena, the terminology is not used consistently.
At high packing fractions, the phenomenology depends crucially on details of the model under consideration because rods are constantly in contact. 
Phase-diagrams can therefore not be mapped one to one~\cite{wensink_emergent_2012,abkenar_collective_2013,weitz_selfpropelled_2015,kuan_hysteresis_2015,shi_self_2018}.

Self-propelled rods have a specific type of interaction with boundaries and obstacles~\cite{wensink_aggregation_2008}:~in contrast to a passive particle where collisions (against a wall or obstacle) are assumed to be instantaneous inducing a reflection, self-propelled rods push persistently against confining walls. 
Consequently, they spend a non-negligible time close to walls and may accumulate there~[Fig.~\ref{fig:phenom:dry_theory}(g)] as observed in granular self-propelled rods~\cite{kudrolli_swarming_2008}.  
However, a torque is induced due to their anisotropic shape implying their alignment with walls; as a result of this, they will slip off collectively, cf.~middle panel of Fig.~\ref{fig:phenom:dry_theory}(g). 
This implies anomalous fluctuations of the force exerted on confining walls; consequently, the mechanical pressure is not a state function in contrast to self-propelled spheres~\cite{solon_pressure_2015}, i.e.~there is no equation of state for self-propelled rods
Another consequence is the nontrivial influence of boundary conditions or spatial inhomogeneities (disordered environments) on the macroscopic emergent patterns. 
The tendency of rods to form clusters and accumulate at walls enables the construction of traps~\cite{kaiser_how_2012,kaiser_capturing_2013} as shown in~[Fig.~\ref{fig:phenom:dry_theory}(h)] or allows using them to propel micro-gears~\cite{kaiser_transport_2014,kaiser_motion_2015}.

The self-propelled rod concept has been extended in several regards: 
\begin{itemize}
	\item[--] Soft deformable particles with an elliptical shape may be considered self-propelled rods, see~\cite{ohta_dynamics_2017} for a recent review. These particles organize themselves into lanes~\cite{menzel_soft_2012} for high aspect ratios, analogous to~Fig.~\ref{fig:phenom:dry_theory}(e). For different parameter values, however, it turned out that the large-scale phenomenology is analogous to the original Vicsek model for polar particles~\cite{vicsek_novel_1995}:~the emergent large-scale order is of polar nature and soliton-like waves were found, as in~\cite{gregoire_onset_2004,chate_collective_2008} for the Vicsek model. 
	\item[--] Flexibility of individual rods, relevant for the description of motility assays~\cite{schaller_polar_2010,weber_random_2015} but also observed in recordings of bacteria~\cite{peruani_collective_2012}, was addressed regarding the individual and collective dynamics in~\cite{duman_collective_2018}. 
	\item[--] The transition towards orientational order in three dimensions was addressed in~\cite{bott_isotropic_2018}.
\end{itemize}

In short, self-propelled rods display a variety of nonequilibrium patterns controlled by shape and density, including polar clusters, lanes, bands, aggregates and topological defects as well as macroscopic orientational order.

\subsubsection{Strong self-propulsion~--~alignment-based description of self-propelled rods}
\label{sec:vic-type_rods}

We now turn to another realm of dry self-propelled rods, namely the limit where self-propulsion is stronger than repulsion, thereby giving rise to quasi two-dimensional dynamics. 
In this limit, the force term $\vec{f}_2$ in Eq.~\ref{eqn:mot} can be neglected, but not the interaction torque, which, as explained below, plays an essential role in the emerging collective patterns.
Though the (aligning) torque should be deduced from a specific interaction mechanism~(Fig.~\ref{fig:dry_hard_models}) strictly speaking, it may heuristically be simplified recalling Onsager's argument~\cite{onsager_effects_1949} regarding its uniaxial~(nematic) symmetry~\cite{peruani_mean_2008}:
% % 
% \begin{align}
% \label{eqn:mot:part}
% 	\dot{\vec{r}}_j(t) &\simeq v_0 \vec{e}_{\parallel} \! \left[ \varphi_j(t) \right] \!, \; \:
% 	\dot{\varphi}_j(t) \simeq \gamma \sum_{k=1}^N \beta \! \left( \vec{r}_{k} - \vec{r}_j  \right) \sin \! \Big[ 2 \big ( \varphi_{k}(t) - \varphi_{j}(t) \big ) \Big ] \!\!\:+\!\!\: \sqrt{2 D_{\varphi} } \!\; \xi_j \! \left( t \right) \! . 
% \end{align}
% %
%
\begin{subequations}
\label{eqn:mot:part}
\begin{align}
	\dot{\vec{r}}_j(t) &\simeq v_0 \vec{e}_{\parallel} \! \left[ \varphi_j(t) \right] \!, \; \: \\
	\dot{\varphi}_j(t) &\simeq \gamma \sum_{k=1}^N \beta \! \left( \vec{r}_{k} - \vec{r}_j  \right) \sin \! \Big[ 2 \big ( \varphi_{k}(t) - \varphi_{j}(t) \big ) \Big ] \!\!\:+\!\!\: \sqrt{2 D_{\varphi} } \!\; \xi_j \! \left( t \right) \! . 
\end{align}
\end{subequations}
Mathematically speaking, the full dynamics~(Eq.~\ref{eqn:mot}) is reduced to an effective point-particle model with nematic alignment. 
The sine-coupling is the first nontrivial mode of a Fourier expansion of the torque with respect to the rod orientation\footnote{Recent work has, however, identified additional torques to be relevant that couple the relative position of rods with their orientation~\cite{grossmann_particle_2019}. }. 
There exists a variety of real-world rod systems with quasi two-dimensional dynamics for which alignment-based descriptions are adequate, for example motility assays~\cite{sumino_large-scale_2012,huber_emergence_2018} or filamentous bacteria confined to move within a thin fluid layer~\cite{nishiguchi_long_2017}.

\begin{figure}[t]
\begin{center}
\includegraphics[width=\textwidth]{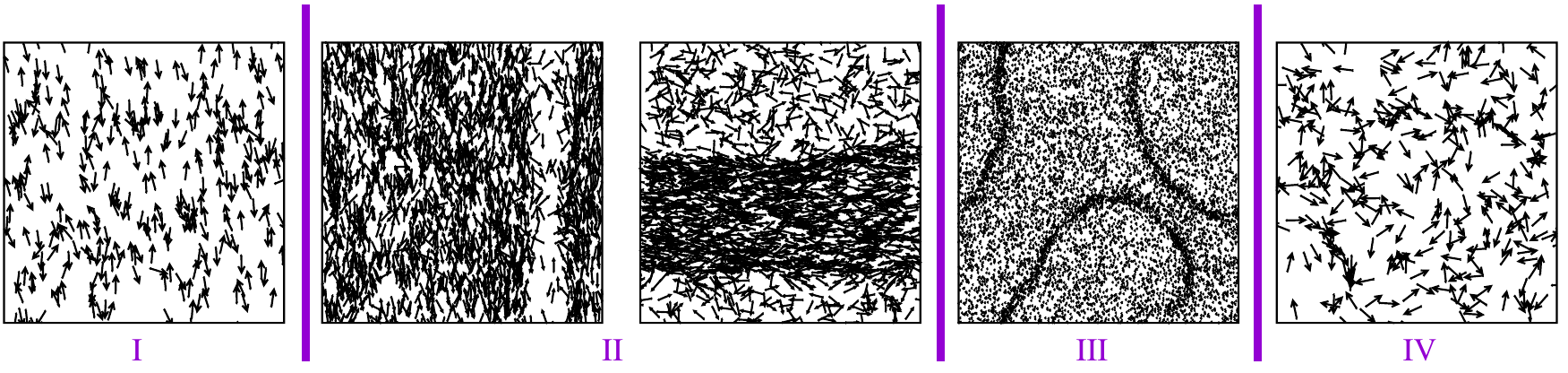}
\end{center}
\vspace{-0.4cm}
\caption{Phenomenology displayed by self-propelled rods with nematic alignment interactions (reproduced from Ginelli~\textit{et~al.}~\cite{ginelli_large_2010}): (I) fluctuating nematic, phase; (II) phase-segregated state composed of a nematic band in a disordered gas; (III) transversally unstable nematic band; (IV) disordered, spatially homogeneous state. }
\label{fig:pheno_vic_rods}
\end{figure}

The phenomenology displayed by Eq.~\ref{eqn:mot:part}~(Fig.~\ref{fig:pheno_vic_rods}) does not depend on the specific functional form of the torque, but on its symmetry~\cite{ginelli_large_2010,shi_self_2018,grossmann_particle_2019}.  
We are reviewing an implementation of the dynamics in the spirit of the seminal Vicsek model~\cite{vicsek_novel_1995}, introduced in~\cite{chate_modeling_2008,ginelli_large_2010}

For low noise or high particle density, the system orders nematically at the scale of the simulation box~(phase I).
In this phase, density fluctuations are anomalously high:~the standard deviation~$\sigma_n$ of the particle number in a subsystem scales with the corresponding mean~$\mean{n}$ as~$\sigma_n \sim \mean{n}^{\alpha}$ with $\alpha > 1/2$ in contrast to~$\alpha = 1/2$ in equilibrium systems. 
These so-called \textit{giant number fluctuations} are a hallmark of phases with orientational order in active matter which were first pointed out in the context of the Vicsek model for collective motion of polar particles~\cite{toner_flocks_1998}. 
In numerical simulations of self-propelled rods,~$\alpha \approx 0.8$ was inferred~\cite{ginelli_large_2010}.
A fluctuating nematic phase was observed in filamentous, non-tumbling mutants of \textit{E.~coli}~\cite{nishiguchi_long_2017}, where~$\alpha \approx 0.6$ was measured.

The question whether nematic order is of long-range nature, i.e.~whether there is a finite level of nematic order in the thermodynamic limit, has not been fully answered yet. 
Recall that continuous symmetry breaking in corresponding equilibrium systems, e.g.~the XY-model, leads to quasi long-range order via a Berezinskii-Kosterlitz-Thouless transition, but long-range order is absent~\cite{berezinskii_destruction_1971,kosterlitz_ordering_1973,mermin_absense_1966}. 
A finite size scaling within particle-based simulations of self-propelled rods suggest indeed the emergence of true long-range order, however, the accessible system sizes are of the order of the persistence length of particle trajectories within the nematic phase~\cite{ginelli_large_2010}. 
A conclusive stochastic field theory for self-propelled rods answering these questions is still missing.

For intermediate fluctuations strengths, rods phase segregate into a high density band with nematic order that is surrounded by a disordered, low density gas~(phase II).
These bands are only observable if the system is large enough. 
This phenomenon is a consequence of the intrinsic coupling of local ordering and density instabilities. 
The width of bands decreases as the noise is increased. 
Thin, elongated, narrow bands undergo a transversal instability giving rise to intermittent dynamics:~global disorder with large fluctuations~(phase III).
Corresponding kinetic theories suggest that the phase-separation process at the mesoscale is analogous to liquid-gas phase separation~\cite{grossmann_mesoscale_2016}, however, bands are always transversally unstable in the thermodynamic limit~\cite{peshkov_nonlinear_2012,grossmann_mesoscale_2016}. 
Recently, another type of band instability leading to its breakup in transversal segments was reported~\cite{cai_dynamical_2019}. 
Finally, a spatially homogeneous, disordered state~(phase IV) is found for high noise levels.

Several extensions of the model class described above were addressed:
\begin{itemize}
	\item[--] Besides nematic order, chiral states were reported~\cite{breier_spontaneuos_2016} for self-propelled rods in three dimensions:~the system organizes into stacked sheets, where nematic order exists in each sheet, but the nematic director rotates from one to another. 
	\item[--] Vortex lattices and active foams can be observed if the white Gaussian noise~(cf.~Eq.~\eqref{eqn:mot:part}) is replaced by an Ornstein-Uhlenbeck process~\cite{nagai_collective_2015}~(correlated noise)~--~a model that explains pattern formation in motility assays~\cite{sumino_large-scale_2012}. 
	\item [--] Trapping and localization phenomena were addressed for the dynamics of self-propelled rods in disordered environments~\cite{peruani_cold_2018}.
\end{itemize}

\subsubsection{Discussion:~effective polar vs.~nematic alignment for dry self-propelled rods}
\label{sec:polvsnem}

Comparing the phenomenology reviewed so far, we note that the \textit{effective alignment interaction} dictating the symmetry of the emergent nonequilibrium patterns in ensembles of self-propelled rods can be either of polar \textit{or} nematic symmetry~--~in contrast to their equilibrium counterparts~--~depending on the strength of self-propulsion, though the microscopic interactions are strictly \textit{uniaxial in all cases}:~if repulsion dominates, colliding clusters of rods may impede each others motion, rods slow down and the clusters fragment or fuse into one single polar cluster; if self-propulsion dominates or if rods can slide over each other, however, they simply walk through each other. 
Penetrable self-propelled rods as studied in~\cite{abkenar_collective_2013} which do repel each other via a soft-core potential are a intermediate case.  
Only recently, the transition from self-propelled rods with volume exclusion~(Section~\ref{sec:rod_ve}), on the one hand, and the alignment-dominated regime~(Section~\ref{sec:vic-type_rods}), on the other hand~\cite{shi_self_2018,grossmann_particle_2019} were studied. 
Notably, a bistable regime where polar structures coexist with nematic bands was recently reported in~\cite{grossmann_particle_2019}, similar to the observed polar-nematic coexistence in motility assays~\cite{huber_emergence_2018} where, moreover, the effective alignment was measured by analyzing binary collisions~\cite{huber_emergence_2018}. 
Intermittent switching between polar and nematic bands was observed before in Vicsek-type models with competing polar- and nematic alignment~\cite{ngo_competing_2012}. 
In short, uniaxially invariant interactions of self-propelled rods do not exclude the emergence of macroscopic polar order.

\subsection{Kinetic theories for self-propelled rods}
\label{sec:kin:theo:dry}

Considerable work has been invested in deriving kinetic descriptions for ordering~\cite{peruani_mean_2008} or clustering~\cite{peruani_nonequilibrium_2006,peruani_cluster_2010,peruani_kinetic_2013} as well as hydrodynamic theories~\cite{,baskaran_enhanced_2008,baskaran_hydrodynamics_2008,baskaran_noneq_2010,harvey_continuum_2013,peshkov_nonlinear_2012,bertin_comparison_2015,degond_continuum_2015,grossmann_mesoscale_2016,grossmann_particle_2019} for self-propelled rods to be reviewed in this section.  
The term \textit{hydrodynamic} refers in this context to the large-scale dynamics on slow time-scales and should not be confused with the long-ranged interactions via a solvent, which is the topic of section~\ref{sec:wetSPR}.

\subsubsection{Fragmentation-coagulation equations for clustering of self-propelled rods}
\label{sec:clu_the}

A kinetic theory for clustering of self-propelled rods was proposed in~\cite{peruani_nonequilibrium_2006,peruani_cluster_2010,peruani_kinetic_2013} which considers clusters to be quasi-particles that may (i) fuse upon collision or (ii) loose single rods at their boundary stochastically.
Formally, these coagulation-fragmentation processes can be considered \textit{reactions} of the form
\begin{align}
	C_k + C_j \, \xrightarrow{\;c_{j,k}\;} \; C_{k + j}, \quad C_{k} \, \xrightarrow{\;f_{k}\;} \; C_{k-1} + C_1,
\end{align}
where~$C_k$ denotes a cluster containing~$k$ rods. 
The fragmentation rate~$f_k = k^{\beta} / \tau$ is determined by a characteristic time~$\tau$ during which single rods stay at the boundary of a cluster and the number of rods at the boundary. 
The parameter~$\beta$ accounts for cluster shape:~$\beta=1/2$ corresponds to circular clusters, whereas~$\beta \simeq 1$ is justified for elongated ones.
The coagulation rate~$c_{j,k} \propto q_s \sigma_0 \left( j^{\alpha} + k^{\alpha}  \right) v_c / L^2$, i.e.~the collision rate of clusters~$C_j$ and $C_k$, is given by the probability~$q_s$ that a collision event leads to successful fusion, the scattering cross section of an individual rod~$\sigma_0$, the size and shape of the respective clusters, the characteristic speed~$v_c$ at which polar clusters move~$v_c$~--~approximately equal to the speed of individual rods~--~and the cluster density.

\begin{figure}[t]
\begin{center}
\includegraphics[width=\textwidth]{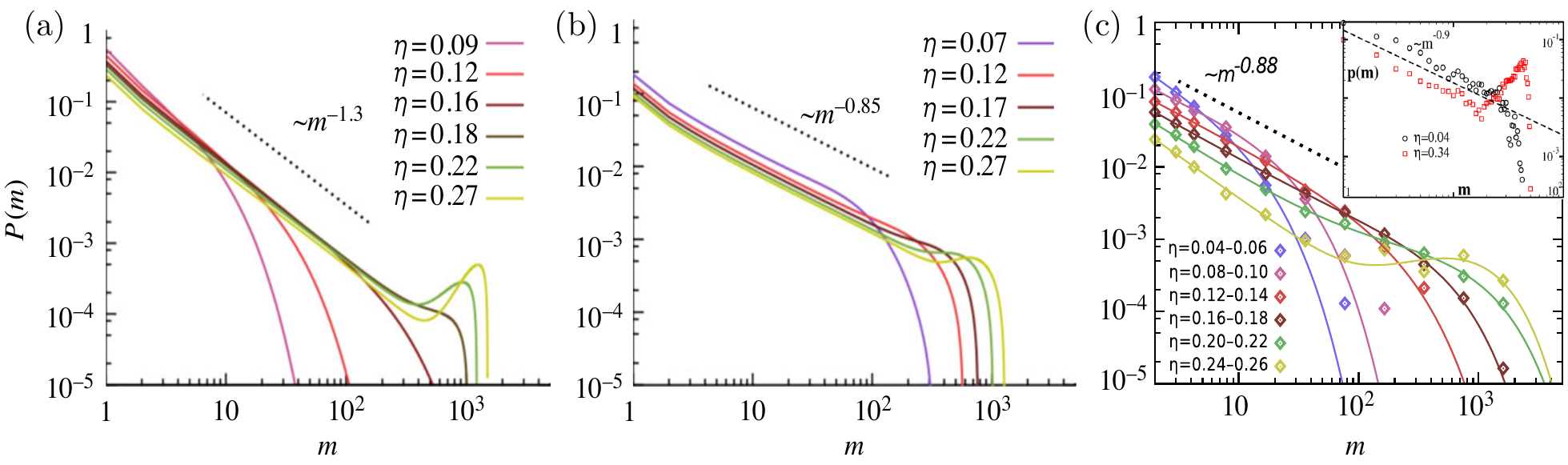}
\end{center}
\vspace{-0.3cm}
\caption{(a-b)~Cluster size distributions~$P(m)$, predicted by the kinetic theory~(Eqs.~\ref{eqn:sm_cl:the}), reprinted from Starru{\ss} \textit{et al.}~\cite{starruss_pattern_2012}. The scaling of the cluster size distribution at the transition depends on the shape of clusters~\cite{peruani_kinetic_2013}: (a)~round clusters~($\beta = 0.5$); (b)~elongated clusters~($\beta = 1$). In each panel, the clustering transition as a function of the packing fraction~$\eta$ is shown. On the right, the distribution found experimentally for non-reversing mutants of \textit{Myxococcus xanthus} is shown (from Peruani~\textit{et~al.}~\cite{peruani_collective_2012}). The inset illustrates a similar transition found earlier in simulations of self-propelled rods~\cite{peruani_nonequilibrium_2006}. }
\label{fig:dry_clustering}
\end{figure}

Assuming that the system is well-mixed and globally disordered\footnote{Clustering in phases with orientational order is discussed in~\cite{huepe_itermittency_2004,peruani_cluster_2010,peruani_kinetic_2013}. } such that~(spatial) correlations are irrelevant, the rate equations $\!\!\!\!$
\begin{align}
\label{eqn:sm_cl:the}
\!\!\dot{n}_j(t) = \left\{ 
	\begin{array}{ll} 
		\phantom{-} 2 f_2 n_2 + \sum_{k=3}^N f_k n_k - \sum_{k=1}^{N-1} c_{k,1} n_k n_1, &  j = 1, \\
		& \\
		- f_{N} n_{N} + \frac{1}{2} \sum_{k=1}^{N-1} c_{k,N-k} n_{k} n_{N-k}, & j = N, \\
		& \\
		\phantom{-}f_{j+1} n_{j+1} - f_{j} n_{j} - \sum_{k=1}^{N-j} c_{k,j} n_k n_j + \frac{1}{2} \sum_{k=1}^{j-1} c_{k,j-k} n_{k} n_{j-k}, \; & \mbox{else} \\
	\end{array} 
	 \right. 
\end{align}
describe mean number~$n_j(t)$ of $C_j$-clusters~\cite{peruani_nonequilibrium_2006}. 
In the homogeneous phase, the cluster size distribution~$P(k) = n_k k / N$~--~the probability of finding a rod in a cluster of type~$C_k$~--~decays exponentially whereas the emergence of a second peak~(bimodal distribution) signals clustering~(Fig.~\ref{fig:dry_clustering}). 
At the transition,~$P(k)$ decays like a power law if~$\alpha = \beta$, and the corresponding exponent was reported to depend on the underlying cluster shape via the exponent~$\beta$~\cite{peruani_kinetic_2013}. 
The fragmentation-coagulation equations can reproduce the clustering transition observed in mutants of the soil bacterium \textit{Myxococcus xanthus}~\cite{peruani_collective_2012,starruss_pattern_2012}.

\subsubsection{Hydrodynamic field theories for dry self-propelled rods}
\label{sec:kin:theo:d}

Now we turn to continuum field theories derived from the microscopic dynamics of self-propelled rods~(Eq.~\ref{eqn:mot}) that provide the basis for identifying phases of self-propelled rods, the linear stability of patterns, associated phase transitions and a characterization of the full nonlinear dynamics at the macroscale. 
Novel theoretical concepts to describe the observed nonequilibrium phenomenology are needed because these systems, by virtue of being driven, evade the constraints of fundamental physical laws from equilibrium systems, like the conservation of energy or momentum, the \textit{action-reaction} principle or fluctuation-dissipation relations.

\paragraph{The objective of kinetic theory}

A typical way to construct continuum equations from particle-based models or observations is the definition of the microscopic particle density\footnote{In this section, we restrict the discussion to two-dimensional systems; two- and three-dimensional systems are examined in Section~\ref{sec:wetSPR}. }
\begin{align}
	\mathcal{P} \! \left( \vec{r},\varphi,t \right) = \sum_{j=1}^N \delta \big( \vec{r} - \vec{r}_j(t) \big) \!\: \delta \big ( \varphi - \varphi_j(t) \big ) 
	\label{eqn:mic:pda}
\end{align}
and its ensemble averaged analogue~$p \! \left( \vec{r},\varphi,t \right) = \mean{\mathcal{P} \! \left( \vec{r},\varphi,t \right)}$. 
The sum over $\delta$-function basically counts how many rods there are at position~$\vec{r}$ with an orientation~$\varphi$. 
Given these distributions, all order parameters of interest could be calculated, such as the particle density~$\rho \! \left( \vec{r},t \right)$, as well as fields for the level of local polar and nematic order, $\vec{P} \! \left( \vec{r},t \right)$ and~$\mathcal{Q} \! \left( \vec{r},t \right)$, respectively, via 
\begin{align}
	\rho    &= \overline{1}, \quad 
	\rho  \vec{P}  = \overline{ \vec{e}_{\parallel} \!\!\: \! \left [ \varphi \right ] },  \quad 
	\left \{ \rho \mathcal{Q}\right \}_{\mu\nu}  =  \overline{2 \, e_{\parallel ,\mu} \!\!\: \! \left [ \varphi \right ] e_{\parallel ,\nu} \!\!\: \! \left [ \varphi \right ]  - \delta_{\mu\nu} }
\end{align}
with~$\overline{f(\varphi)} = \int_{-\pi}^{\pi} d \varphi \, f(\varphi) p \! \left( \vec{r},\varphi,t \right)$. 
In general, all the three order parameters are relevant as polar and/or nematic order are observed in ensembles of self-propelled rods, and the emergence of order is inherently related to density instabilities. 
Note that the ansatz to describe the large-scale dynamics of self-propelled rods via order-parameter fields differs fundamentally from the complementary approach based on cluster statistics discussed in Section~\ref{sec:clu_the}; a hybrid approach combining elements of both types of theories was suggested in~\cite{harvey_continuum_2013} to describe clustering of myxobacteria qualitatively.

\paragraph{Variants of kinetic theories}

There are several ways to construct kinetic theories:~one may propose or derive an equation for the fluctuating field~$\mathcal{P} \! \left( \vec{r},\varphi,t \right)$ directly~\cite{dean_langevin_1996}, or perform some temporal, spatial or ensemble-averaging procedure. 
The type of coarse-graining determines whether the resulting partial differential equations are stochastic or deterministic~\cite{archer_dynamical_2004}. 
Kinetic theories for self-propelled rods were mostly build upon a Fokker-Planck~(Smoluchowski)~\cite{peruani_mean_2008,baskaran_enhanced_2008,baskaran_hydrodynamics_2008,baskaran_noneq_2010,degond_continuum_2015,grossmann_mesoscale_2016,grossmann_particle_2019} or a generalized Boltzmann approach~\cite{peshkov_nonlinear_2012,bertin_comparison_2015}.

A direct way to obtain a kinetic equation for the fluctuating particle distribution~$\mathcal{P} \! \left( \vec{r},\varphi,t \right)$ is the application of stochastic calculus rules to the change of variables given by Eq.~\ref{eqn:mic:pda} yielding~\cite{dean_langevin_1996}
\begin{subequations}
\label{eqn:ThedActMat}
\begin{align}
\label{eqn:mpda}
	\partial_t \mathcal{P} + v_0 \vec{e}_{\parallel} \!\!\: \! \left [ \varphi \right ] \!\!\: \cdot \!\!\: \nabla \mathcal{P} = D_{\varphi} \partial_{\varphi}^2 \mathcal{P} 
	- \!\!\: \nabla \!\!\: \cdot \!\!\: \Big \{ \mathcal{P} \vec{F} \! \left[ \mathcal{P} \right] \! \Big \} \!\!\:
	- \!\!\: \partial_{\varphi} \Big \{ \mathcal{P} M \! \left[ \mathcal{P} \right] \! \Big \} \!\!\:
	+ \!\!\: \nabla \! \!\: \cdot \!\!\: \Big\{ \mathcal{D} \! \left[ \varphi \right] \!\!\: \cdot \!\!\: \nabla \mathcal{P} \Big \} 
	+ \!\, \Xi \! \left[ \mathcal{P} \right] \! . 
\end{align}
The rod-rod interaction turns this equation into a nonlinear integro-partial differential equation:
\begin{align}
\label{eqn:int_op}
		\!\!\! \vec{F} \! \left[ \mathcal{P} \right] \! &= \!\! \!\!\: \int \!\! d^2 r' \!\!\: d\varphi' \mathcal{P} \! \left( \vec{r}'\!,\varphi'\!,t \right) \! \vec{f}_2 \! \left( \vec{r}' \! - \vec{r}; \varphi' \!\:\!, \varphi \right) \!, \; \\
		M \! \left[ \mathcal{P} \right] \! &= \!\! \!\!\: \int \!\! d^2 r' \!\!\: d\varphi' \mathcal{P} \! \left( \vec{r}'\!,\varphi'\!,t \right) \! m_2 \! \left( \vec{r}' \! - \vec{r}; \varphi' \!\:\!, \varphi \right) \! .
\end{align}
\end{subequations}
Formally, this equation is exact. 
It involves, however, a noise~$\Xi \! \left[ \mathcal{P} \right]$ which is  difficult to handle due to its multiplicative nature. 
Macroscopic noise terms were shown to be of importance to address questions such as pattern selection~\cite{solon_from_2015} and the stability of orientationally ordered phases in related nonequilibrium systems~\cite{toner_long_1995,grossmann_superdiffusion_2016}, however, the direct application of Eq.~\ref{eqn:mpda} remains a challenge to self-propelled rods.

\paragraph*{Fokker-Planck approach}

A nonlinear Fokker-Planck equation is obtained by ensemble-averaging Eq.~\ref{eqn:mpda} or, equivalently, as a marginal probability of the $N$-particle distribution function~\cite{risken_fokker-planck_1996} corresponding to the Langevin dynamics~(Eq.~\ref{eqn:mot}), similar to the BBGKY hierarchy for Hamiltonian systems~\cite{kardar_particles_2007}. 
The main complication is that an appropriate closure relation is needed:~upon averaging, the products~$\mean{\mathcal{P} \! \left( \vec{r},\varphi,t \right) \mathcal{P} \! \left( \vec{r}'\!\!\:,\varphi'\!\!\:,t \right)}$ represent the pair-correlation function which is, in turn, coupled to the three-particle distribution etc.
Commonly, the factorization~$\mean{\mathcal{P} \! \left( \vec{r},\varphi,t \right) \mathcal{P} \! \left( \vec{r}'\!\!\:,\varphi'\!\!\:,t \right)} \approx \mean{ \mathcal{P} \! \left( \vec{r},\varphi,t \right)} \mean{ \mathcal{P} \! \left( \vec{r}'\!\!\:,\varphi'\!\!\:,t \right) }$
corresponding to a \textit{mean-field} or \textit{molecular chaos} approximation is assumed~\cite{peruani_mean_2008,baskaran_hydrodynamics_2008,baskaran_noneq_2010,degond_continuum_2015,grossmann_mesoscale_2016}. 
This ansatz may be justified if the system is well-mixed and one rod interacts with many neighbors, such that fluctuations are rendered irrelevant.

In the context of self-propelled rods pushing each other as described in Section~\ref{sec:rod_ve}, mean-field arguments become less reliable as distinct correlations emerge if rods block each others motion. 
A work-around is the derivation of order-parameter equations where transport coefficients depend on integrals over the pair-correlation function to be measured numerically~\cite{bialke_microscopic_2013,grossmann_particle_2019}. 
The prediction of pair-correlations from first principles, as it was done for self-propelled discs~\cite{haertel_three_2018}, is an open theoretical challenge.

\paragraph*{Boltzmann approach}

Complementary to the Fokker-Planck ansatz, generalized Boltzmann equations were proposed for alignment-based active particles~\cite{peshkov_boltzmann_2014}, and so for self-propelled rods with nematic alignment~\cite{peshkov_nonlinear_2012}, originally developed in the context of the classical Vicsek model~\cite{vicsek_novel_1995,bertin_boltzmann_2006}. 
The basic structure of the Boltzmann equation is analogous to the kinetic theories discussed before, however, the integral operators~(Eq.~\ref{eqn:int_op}) describing collisions are constructed in a different way, based on two physical elements:~(i) what happens if rods collide and~(ii) the collision frequency. 
The latter is a complicated object that depends on particle density, the structure of the system~(clustered or non-clustered), particle speed etc.
The Boltzmann approach relies on the molecular chaos assumption, i.e.~the negligence of pre-collisional angles.

\subsubsection{Predictions, results and open questions}
\label{subs:res:kt}

We believe that having an all-encompassing understanding of dry active matter is tantamount to solving Eq.~\ref{eqn:ThedActMat} since self-propelled discs and dry active nematics are certain limiting cases of dry self-propelled rods~(cf.~Table~\ref{tab:conf_dry2}). 
Up to now, only some limiting cases are well understood.

\paragraph*{Spherical particles}

Torques are absent implying~$M[\mathcal{P}] = 0$ for self-propelled discs. 
In this limit, the full theory is reduced to well-studied field theories for MIPS~\cite{cates_mips_2015}.
Notably, the understanding of MIPS requires to take inter-particles correlations into account~\cite{bialke_microscopic_2013,haertel_three_2018}. 
Some equilibrium concepts like pressure or Maxwell constructions are applicable in this case~\cite{solon_generalized2_2018}.

\paragraph*{Alignment-based models for rod-shaped objects}

If self-propelled rods are soft or the self-propulsion is strong, steric hindering is rendered irrelevant implying~$\mathcal{F} \! \left[ \mathcal{P} \right] \approx 0$ in Eq.~\ref{eqn:int_op}. 
Alignment-based models are applicable in this regime, as discussed in Section~\ref{sec:dry_pheomnea}, which are analytically tractable.

Based on a spatially homogeneous Fokker-Planck approach, the isotropic-nematic mean-field transition for self-propelled rods interacting by velocity-alignment was analyzed in~\cite{peruani_mean_2008} at the level of the distribution function. 
Subsequent works deduced predictions from order parameter equations~\cite{baskaran_enhanced_2008,baskaran_hydrodynamics_2008,baskaran_noneq_2010,peshkov_nonlinear_2012,bertin_comparison_2015,grossmann_mesoscale_2016,grossmann_particle_2019} that are non-linearly coupled among themselves and, moreover, to higher order moments.
Therefore, another hierarchy problem must be overcome by appropriate closure relations. 
The type of decoupling determines nonlinear terms in the resulting field equations~\cite{bertin_comparison_2015}. 
There is no unique way to obtain closed sets of equations, but the type of closure depends on the parameter regime under consideration.

An essential advance was the derivation of the nonlinear field equations~\cite{peshkov_nonlinear_2012,bertin_comparison_2015}
\begin{align}
	\partial_t \rho & \! = 
	- v_0 \nabla \!\!\: \cdot \!\!\: \bar{\vec{P}} , \nonumber \\
	\partial_t \bar{\vec{P}} &\!=\! 
	- \! \left[ \alpha \!\!\:-\!\!\: \frac{\beta}{2} \mbox{tr} \! \left( \bar{\mathcal{Q}}^2 \right) \! \right] \! \bar{\vec{P}}
	- \frac{v_0}{2} \!\!\: \Big( \nabla \rho + \! \nabla \!\!\: \cdot \!\!\: \bar{\mathcal{Q}} \Big) 
	+ \frac{\gamma}{2} \left ( \bar{\mathcal{Q}} \!\!\:\cdot \!\!\: \nabla  \right ) \!\!\:\cdot \!\!\: \bar{\mathcal{Q}} 
	+ \zeta \!\: \bar{\vec{P}} \cdot \bar{\mathcal{Q}} , \\
	\partial_t \bar{\mathcal{Q}} &\!=\! \!
	\left[ \mu \! - \! \frac{\xi}{2}  \mbox{tr} \! \left( \bar{\mathcal{Q}}^2 \right) \! \right] \!\!\:\!\bar{ \mathcal{Q}} 
	\!-\! \frac{v_0}{2} \left ( \nabla \bar{\vec{P}} \right )^\text{st}  \!\!\:\!\!\:
	\!+\! \frac{\nu}{4} \Delta \bar{\mathcal{Q}} 
	\!+\! \omega \left ( \bar{\vec{P}} \bar{\vec{P}} \right )^\text{st} \!\!\: \!\!\:
	\! +\! \tau \! \abs{\bar{\vec{P}}}^2 \! \bar{\mathcal{Q}} 
	\!-\! \frac{\tilde{\chi}}{2} \nabla \!\cdot \!\left ( \bar{\vec{P}} \bar{\mathcal{Q}} \right )^\text{st} \!\!\:\!\!\:
	\!-\! \frac{\tilde{\kappa}}{2} \bar{\vec{P}} \! \cdot \! \left ( \nabla  \bar{\mathcal{Q}} \right )^\text{st}\!\!\: \nonumber
\end{align}
for~$\bar{\vec{P}} = \rho \vec{P}$ and $\bar{\mathcal{Q}} = \rho \mathcal{Q}$ which predicts the phenomenology shown in Fig.~\ref{fig:pheno_vic_rods}; the shape of bands was, notably, analytically obtained~\cite{peshkov_nonlinear_2012}. 
As the mesoscale patterns are nematic on the local and global scales, the relevance of the polar order parameter for this type of point-particle system with alignment has been questioned~\cite{grossmann_mesoscale_2016}~--~the shape of bands, their transversal instability in the thermodynamic limit~\cite{ngo_large_2014,grossmann_mesoscale_2016} and the phase diagram for finite-sized systems~\cite{grossmann_mesoscale_2016} are also obtained from a reduced set of equations for density and nematic order parameter, derived via the Fokker-Planck approach. 
An in-depth comparison of the Boltzmann and Fokker-Planck approach for self-propelled rods can be found in~\cite{bertin_comparison_2015}.
The theoretical question whether the fluctuating nematic phase is stable in the thermodynamic limit, i.e.~whether true long-range exists in this system, is, however, unanswered.

Self-propelled rods with velocity reversal were analyzed in one dimension with regard to bacterial self-organization~\cite{gejji_macroscopic_2012}; their relation to dry active nematics and the observed defect dynamics was addressed~\cite{shi_topological_2013,shi_instabilities_2014} and their mesoscale pattern formation was reconstructed from numerical continuation of the underlying continuum theory~\cite{grossmann_mesoscale_2016}.

\paragraph{Outlook}
\label{sec:dryoutlook}

The full phenomenology displayed by self-propelled rods with anisotropic repulsive interactions and aligning torque~(Fig.~\ref{fig:phenom:dry_theory}) are less well understood. 
Particularly the explanation of large-scale polar structures remains a theoretical challenge as correlations~--~beyond mean-field or, equivalently, molecular chaos approximations~--~need to be taken into account. 
Early theoretical works considered self-propelled rods within generalized Onsager theories~\cite{baskaran_enhanced_2008,baskaran_hydrodynamics_2008,baskaran_noneq_2010}:~an enhanced tendency towards nematic ordering due to activity~(shift of the critical point of the isotropic-nematic transition known from passive rods), novel types of instabilities in the nematic phase and the crucial influence of boundaries were predicted. 
Further, the importance of a polar order parameter field has been pointed out, however, states with macroscopic polar order were not expected to emerge~\cite{baskaran_hydrodynamics_2008} as the dynamics of the polar order parameter was found to be fast.
Recently, a semi-analytical theory which takes orientational correlations into account suggested, in contrast, that the polar order parameter may become unstable at the spatially homogeneous level and may, thus, turn into a slow hydrodynamic variable~\cite{grossmann_particle_2019}. 
At the moment, there is no conclusive theory that accounts for all phenomena displayed by self-propelled rods, including laning, polar clustering, polar bands, giant clusters and nontrivial interaction with boundaries of the system. 
In particular, the understanding jamming and glassy dynamics at high densities~--~where rods are constantly in contact and microscopic details of the interaction mechanisms matter consequently, remains a theoretical challenge.

%%%%%%%%%%%%%%%%%%%%%%%%%%%%%%%%%%%%%%%%%%%%%%%%%%%%%%%
%%%%%%%%%%%%%%%%%%%%%%%%%%%%%%%%%%%%%%%%%%%%%%%%%%%%%%%

\section{WET SELF-PROPELLED RODS}
\label{sec:wetSPR}

\begin{figure}[t]
    \centering
    \includegraphics[width=\textwidth]{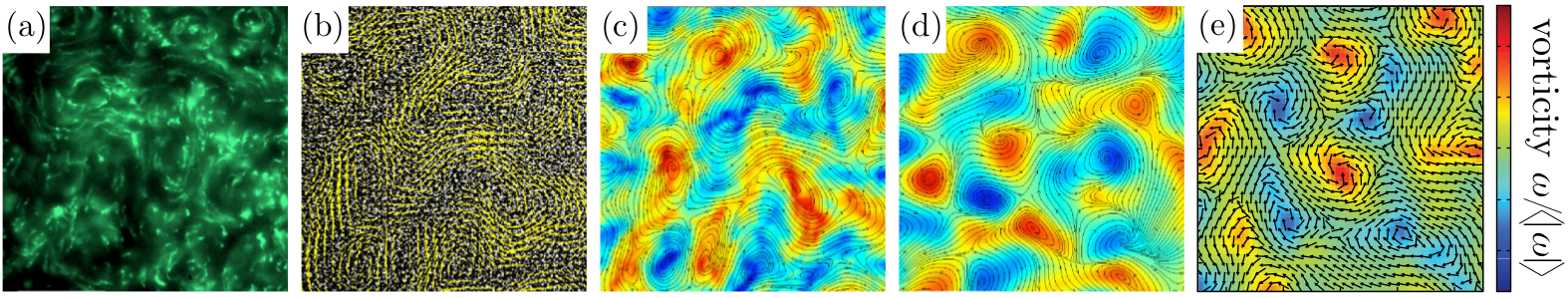}
    \vspace*{-0.4cm}
    \caption{Flow orientation fields extracted from experiments and simulations~(2D slice from a 3D system). A:~turbulent of fluorescent tracer particles in suspension of \emph{B.~subtilis}; B:~suspension overlaid with PIV flow fields; C:~streamlines and normalized vorticity field; D:~snapshot of continuum simulations~(A-D from Dunkel \textit{et al.}~\cite{dunkel_fluid_2013}); E:~snapshot of a 2D particle simulation~\cite{grossmann_vortex_2014}. }
    \label{CompTurb}
\end{figure}

The dynamics of microswimmers, e.g.~swimming bacteria, is strongly influenced by long-ranged hydrodynamic interactions. 
Hydrodynamics together with the incompressibility of the flow changes fundamentally the physics of the problem leading to novel emergent states, including irregular, seemingly turbulent vortex dynamics and topological defects with an intrinsic length scale~--~the characteristics of mesoscale turbulence~\cite{wensink_meso-scale_2012} as illustrated in~Fig.~\ref{CompTurb} for a three dimensional bacterial suspensions~\cite{dunkel_fluid_2013}. 
Hence, a model of self-propelled swimming rods has to include the surrounding fluid and its impact on the dynamics of rods. 
Henceforth, we refer to this type of systems as \emph{wet self-propelled rods}.

We saw in the previous section that contact-mediated interactions of dry self-propelled rods effectively leads to polar or nematic alignment. 
That is why we focus in this section on the interplay and competition between either polar or nematic near-field alignment of active swimming particles and hydrodynamic coupling via a solvent. 
Using simplified alignment-based interactions enables the derivation of coarse-grained continuum equations that describe the large-scale behavior of wet self-propelled rods, which are reviewed in this section. 
We will concentrate on the ability of different models of wet self-propelled rods to reproduce mesoscale turbulence.

\subsection{Nonequilibrium dynamics of wet self-propelled rods}

In contrast to fishes, mussels and humans, the motion of microswimmers is dominated by viscous dissipation and the absence of inertia~--~the Reynolds number of the flow is very small~\cite{purcell_life_1977}.
The velocity field $\vec{u} \! \left( \vec{r},t \right)$ generated by a microswimmer is thus approximately described by the Stokes equation in conjunction with the incompressibility condition
\begin{equation}
\label{StokesEq}
     \mu_0 \Delta {\bf u} = \nabla p - \nabla \cdot \boldsymbol{\sigma},\; \; \nabla\cdot {\bf u} = 0. 
\end{equation}
Recently, the far-field flow around motile bacteria and micro-algae were measured~\cite{drescher_fluid_2011,drescher_direct_2010}. 
Both studies reported that the fluid flow far from the swimmer is well approximated by a force dipole. 
Therefore, we consider wet self-propelled rods as elongated objects suspended in a fluid, equipped with a force dipole~\cite{lauga_hydrodynamics_2009,elgeti_physics_2015,saintillan_rheology_2018}.

A collection of wet self-propelled rods is characterized by short-range contact and long-range hydrodynamic interactions mediated by the solvent fluid.
Short range interactions of microswimmers are very complex due to elastic deformations, lubrication effects, chemical signaling and near field hydrodynamics. 
For an effective alignment-based description of these pair interactions, one can consider the angle before and after collisions of two self-propelled rods which can have nematic and polar contributions~\cite{huber_emergence_2018}, cf.~the discussion in Section~\ref{sec:polvsnem}: the dynamics of a collection of~$N$ wet self-propelled rods in two or three spatial dimensions is thus described by an extension of the overdamped Langevin Eq.~\ref{eqn:mot} where anisotropic repulsion and torque are recast into an effective velocity-alignment rule and coupling of the rods' orientation~$\vec{e}_{\parallel}$ to the flow~$\vec{u}$ is included.
For convenience, we neglect the anisotropy of diffusion~($D_{\parallel} = D_\perp = D$) in this section~\cite{heidenreich_hydrodynamic_2016}.

The coupling of the orientation to the solvent motion is assumed to be in line by Jeffrey's theory for the oscillatory tumbling motion of elongated passive rods~\cite{jeffery_motion_1922}. 
The coupling to fluid rotations, ${\boldsymbol \Omega} = \left [\nabla {\bf u} - (\nabla {\bf u})^T \right ]/2$ is different from the coupling to fluid strain deformations ${\boldsymbol \Sigma}= \left [\nabla {\bf u} + (\nabla {\bf u})^T\right ]/2$ due to the particle shape given by the aspect ratio~$\chi$. 
The coupling parameter  $a_0=(\chi^2-1)/(\chi^2+1)$ for passive rods is oftentimes assumed~\cite{heidenreich_hydrodynamic_2016}, but it may be different for microswimmers~\cite{rafai_effective_2010}.

The solvent velocity is determined by Eqs.~\ref{StokesEq} with an averaged stress tensor $\boldsymbol{\sigma} $ composed of a passive and an active component: $\boldsymbol{\sigma} = {\boldsymbol\sigma}^p + {\boldsymbol\sigma}^a$. 
The passive stress~${\boldsymbol\sigma}^p$ is also present for non-moving rods: it consists of an isotropic contribution and a part explicitly depending on the rod's orientation. 
The anisotropic part is proportional to the Landau-de Gennes potential from liquid crystal theory~\cite{borgmeyer_unified_1995} and can be approximated by ${\boldsymbol \sigma}^p \approx  p_\mathrm{kin} \vartheta(\rho) \mathcal{Q}$ where $\rho$ is the number density, $p_\mathrm{kin}$ the equilibrium kinetic pressure, $\vartheta(\rho)$ is the coefficient of the first term (lyotropic liquid crystals) and $\mathcal{Q}$ is the nematic order parameter tensor. 
The isotropic part describes the viscosity change of a fluid when spherical colloidal particles are added which depends  strongly on the density of colloids as given by the Batchelor-Einstein equation~\cite{batchelor_transport_1974}: $\mu^* = \mu_0(1+ k_1 \rho + k_3\rho^2)$. 
The description of the active stress~${\boldsymbol\sigma}^a$ goes back to two seminal papers that demonstrated the active stress of wet rods to be proportional to the nematic order parameter with a sign depending on the forcing of the individuals, i.e.~pusher or puller~\cite{simha_statistical_2002,potomkin_focusing_2017,hatwalne_rheology_2004}. 
Recently, the representation of the active stress tensor similarly to the approach used in  \cite{liverpool2008hydrodynamics} was extended by  including higher derivatives~\cite{heidenreich_hydrodynamic_2016} via a Taylor expansion of point forces yielding
\begin{equation}
\label{A:stress_2}
\bs \sigma^a \approx -f_0 \rho \left \{\xi_1 \mathcal{Q} + 2\xi_2 \! \left(\nabla \bP \right)^\text{st}+2\xi_4 \Delta \! \left(\nabla \bP \right)^\text{st} + ... \right \},  
\end{equation}
where $f_0$ is the force of microswimmers onto the surrounding fluid.
The second and third term in the bracket on the r.h.s.~of Eq.~\ref{A:stress_2} describes higher order effects that do not vanish if local polar order is present.
The coefficients $\xi_i$ depend on the length of the microswimmer.

The Langevin dynamics for particles that interact purely by hydrodynamic interactions~(no orientational alignment) was derived from slender body theory in~\cite{saintillan_orientational_2007,saintillan_active_2013} where a kinetic equation for the polar order parameter coupled to the number density and the Stokes equation was obtained~\cite{saintillan_orientational_2007}. 
It was demonstrated that such suspensions are always unstable to fluctuations~\cite{simha_hydrodynamic_2002,saintillan_instabilities_2008b}. 
This approach was later extended to include orientational alignment interactions mimicking steric effects~\cite{ezhilan_instabilities_2013}. 
A related Langevin approach including models for short-ranged interactions was used for simulations~\cite{lushi_modeling_2013}, demonstrating the need to include hydrodynamics to describe experiments in confined suspensions~\cite{lushi_fluid_2014}.

\subsection{Continuum models}

Collective phenomena of wet self-propelled rods can be studied at different scales. 
For investigations of the large scale collective behavior, continuum equations for mesoscopic or macroscopic order parameter fields are often considered just like for dry self-propelled rods~(Section~\ref{sec:kin:theo:dry}). 
To obtain continuum equations for variables like the number density, polar and nematic order parameters, one uses coarse graining as described in Section~\ref{sec:kin:theo:d}. 
As the resulting moment equations are hierarchically coupled, higher moments have to be expressed in terms of lower moments in explicit closure relations to obtain a closed set of equations.
As discussed in the context of dry self-propelled rods, there is no unique closure condition, cf.~\cite{kroger_consistent_2008}, but symmetries of patterns can oftentimes simplify the task. 
In the following, two different symmetry classes are considered for the special case of a high, non-fluctuating swimmer density as suggested experimentally~($\rho\approx\rho_0= \mathrm{const}$).

\subsubsection{Active nematic case}

A generalized quadratic closure relation was proposed in~\cite{reinken_derivation_2018} for situations where ensembles of wet self-propelled rods display local nematic, but no polar, order, which is the case, for example, for certain types of microswimmers~\cite{li_data_2019}. 
In this case, one obtains the following equation for the nematic order parameter~\cite{reinken_derivation_2018}, 
\begin{eqnarray}
\label{nemat_eq}
\p_t{\mathcal{Q}} + \bu \cdot \nabla \!\!\: {\mathcal{Q}} -2 \bs \Omega \cdot {\mathcal{Q}} -  \kappa \left(\bs \Sigma \cdot {\mathcal{Q}}\right)^\mathrm{st} + 2 a_0 \boldsymbol{\Sigma}:(\mathcal{Q} \mathcal{Q})^{st}  =    - \boldsymbol{\Phi}^{\mathcal{Q}} + D \Delta {\mathcal{Q}} + \lambda_K \bs \Sigma ,
\end{eqnarray}
where $\lambda_K$ is the tumbling parameter and $\boldsymbol{\Phi}^{\mathcal{Q}}$ is the derivative of the Landau-de Gennes potential  $\boldsymbol{\Phi} \! \left( \mathcal{Q} \right) = A (\rho) \mbox{tr}[\mathcal{Q}^2 ] + B \mbox{tr}[ \mathcal{Q}^3 ] + C \mbox{tr}[ \mathcal{Q}^2 ]^4 $, well known from liquid crystal theory~\cite{prost_physics_1995}. 
The coefficient $A (\rho)$ can change its sign for different densities, indicating the isotropic-to-nematic phase transition. 
The obtained equations are reminiscent of the order parameter equation of passive liquid crystal subjected to an external flow. 
The activity enters through the flow field $\bf u$ that is generated by an active stress. 
Employing head-tail symmetry to the stress one obtains $\boldsymbol{\sigma}^a = (p_\mathrm{kin}\vartheta (\rho) - \rho f_0 \xi_1) {\mathcal{Q}}$ as a generic model for active nematics~\cite{marchetti_hydrodynamics_2013,cates_shearing_2008}. 
There are several models of active nematics where either the constitutive equation for the stress tensor and the order parameter equation were modified or where the Navier-Stokes instead of the Stokes equation was used~\cite{giomi_geometry_2015,thampi_instabilities_2014}.
Remarkably, all these models for active nematics have a long-wavelength instability of the nematic state~\cite{simha_hydrodynamic_2002} in common that results in a creation, annihilation and motion of topological defects in the nematic order parameter field~\cite{doostmohammadi_stabilization_2016,giomi_defect_2013}.
Consequently, these models explain spatially inhomogeneous states, but they cannot account for the characteristic length-scale found in the mesoscale turbulent state of bacterial suspensions.

\subsubsection{Active polar case}

For suspensions with polar symmetry like certain collections of pushers or pullers, the first nontrivial term in the expansion of the probability distribution function is the polar order parameter~$\vec{P}$.
The following dynamical equation for the polarization was derived using an extended Doi-closure approximation for ${\mathcal{Q}}$ that takes the generation of nematic order by the solvent flow into account~\cite{reinken_derivation_2018},
\begin{equation}
   \label{Res_FieldEq}
(\p_t + \bu\cdot \nabla - \bs \Omega \cdot \bP - \kappa \bs \Sigma \cdot \bP + \lambda_0  \bP \cdot \nabla )\bP =  - \boldsymbol{\Phi} ^{P} - {\nabla} p^*, 
\end{equation}
where $\boldsymbol{\Phi} ^{P}$ is the derivative of a functional  $ \Phi = \alpha \bP^2/2 + \beta  \bP^4 /4 + \Gamma_2 (\nabla  \bP)^2/2 - \Gamma_4 (\nabla \nabla  \bP)^2/2$  representing a vectorial generalization of the scalar potential used for the Swift-Hohenberg equation~\cite{swift_hydrodynamic_1977}. 
A characteristic feature of this potential is that the sign of $\Gamma_2=D + a_1 \rho v_0 - a_2 v_0 f_0/\mu_\mathrm{eff}$ ($a_1>0, a_2>0$) can become negative. Note, that $\Gamma_4$ was found to be always negative \cite{reinken_derivation_2018}.
The contributions of the coefficients related to the polar alignment interaction ($a_1$) and to the hydrodynamics ($a_2$) indicate that large-scale hydrodynamic interactions give rise to a negative sign of $\Gamma_2$ which is significant for the emergence of patterns with a characteristic length-scale that is given by~$\Lambda \simeq \sqrt{\Gamma_4/\Gamma_2}$~\cite{wensink_meso-scale_2012}.

The related active and passive stress contribution was derived as $\mu_\mathrm{eff} \nabla^2 \bu = \boldsymbol{\nabla}\cdot \sigma$, were $\mu_\text{eff} = \mu^* - k_2 \ell f_0\rho$ is the effective bulk viscosity. 
The coefficient $k_2>0$ depends on microscopic details.
Remarkably, the effective viscosity is increased or decreased by the swimmer density $\rho$ depending on the sign of $f_0$. 
Indeed, a corresponding increase of the effective viscosity was observed for pullers~\cite{rafai_effective_2010} and a decrease for pushers~\cite{sokolov_reduction_2009}. 
For intermediate densities the effective viscosity observed can reach  very low values reminiscent of the behavior of superfluids~\cite{lopez_turning_2015}.

\subsubsection{Outlook}

Dry self-propelled rods are a limit case of wet self-propelled rods if the active flow-field generation is absent~($f_0 = 0$) or, alternatively, the fluid flow is neglected. 
In that limit, the coefficient~$\Gamma_2$ in Eq.~\ref{Res_FieldEq} is strictly positive; accordingly, the characteristic instability due to long-ranged hydrodynamics as observed in collections of wet self-propelled rods cannot emerge in their dry analogues~(Section~\ref{sec:drySPR}).

Just like for dry self-propelled rods, a coarse-grained theory derived from the microscopic interactions of wet self-propelled rods including anisotropic repulsion and torque instead of alignment-based descriptions~--~which would require the inclusion of inter-particle correlations beyond the restrictions of mean-field~(molecular chaos) assumptions~--~remains a theoretical challenge for further investigations in the near future~(see discussion in Section~\ref{sec:dryoutlook}).

\subsection{Phenomenology of mesoscale turbulence}

\begin{figure}[t]
    \centering
    \includegraphics[width=\textwidth]{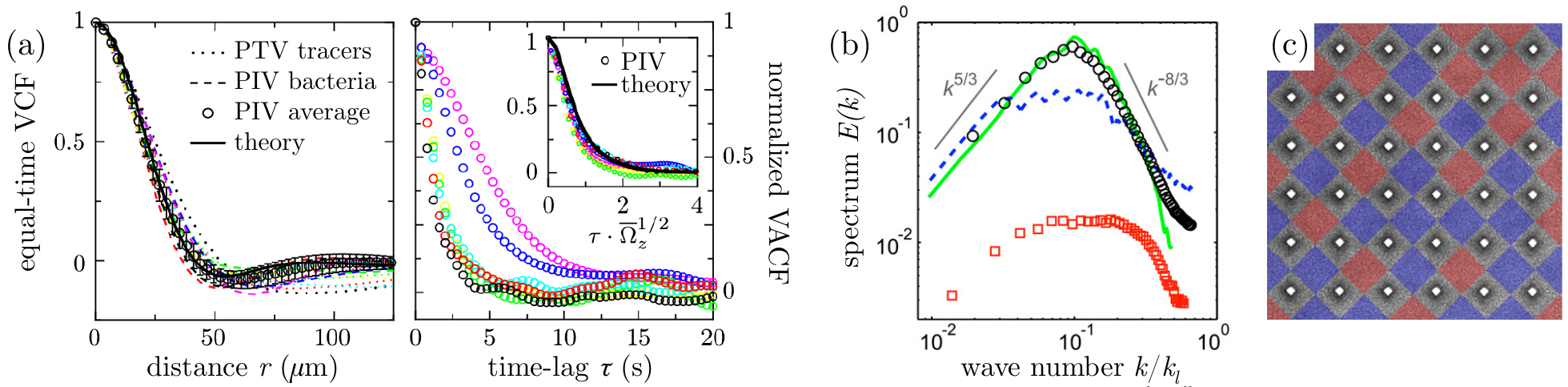}
    \caption{Comparison between continuum theory and experiments. (a) Correlation functions of the solvent (PTV), bacterial flow (PIV) and continuum simulations (theory) for different energies: \textit{left}: equal-time velocity correlation function indicates a characteristic vortex radius; \textit{right}: velocity autocorrelation function collapses when the time lag $\tau$ is rescaled by the enstrophy timescale (inset). The flow memory decreases with increasing activity. (b) 3D spectrum for bulk turbulence is plotted (red squares). Spectra of 2D continuum theory (green line) and quasi-2D experiments (black circles) are in good agreement. (c) Bacterial vortex lattice stabilized by a periodic array of pillars: a snapshot of swimming bacteria overlaid with a color plot indicating the rotational direction of vortices. The emergence of a vortex lattice depends on the lattice constant of the pillar array. Panel (a) adapted from Dunkel~\textit{et~al.}~\cite{dunkel_fluid_2013}, panel (b) adapted from Wensink~\textit{et~al.}~\cite{wensink_meso-scale_2012}, and panel (c) from Nishiguchi~\textit{et~al.}~\cite{nishiguchi_engineering_2018}.}
	\label{lengthScale}
\end{figure}

The self-driven flow of bacterial suspensions has substantially different characteristics from high Reynolds-number Navier-Stokes flow due to the nonequilibrium driving at the microscale. 
The statistical properties of the self-sustained mesoscale turbulence state in active suspensions were first addressed in detail in~\cite{wensink_meso-scale_2012} combing experimental particle image velocimetry~(PIV) analysis, using \textit{Bacillus subtilis} as a model organism, together with the following phenomenological continuum equation which reproduces the experimental observations well: 
\begin{subequations}
\label{GSHE1}
\begin{align}
    \left(\p_{t} +\lambda_0 \vec{P} \cdot \nabla \right) \vec{P} & = - \nabla p  - \alpha  \vec{P} - \beta \abs{\vec{P}}^2 \! \vec{P} + \Gamma_2 \Delta \vec{P} + \Gamma_4  \Delta^2\vec{P}, \; \;  \; \\  
    0 &= \boldsymbol{\nabla} \cdot \vec{P}. 
\end{align}
\end{subequations}
This equation can be considered a generalization of the Swift-Hohenberg equation~\cite{swift_hydrodynamic_1977} for a vectorial orientation field, including an additional nonlinear advection term on the left hand side~\cite{wensink_meso-scale_2012}.
The interrelation of the full hydrodynamic model~[Eqs.~\ref{StokesEq}, \ref{A:stress_2} and~\ref{Res_FieldEq}] and Eq.~\eqref{GSHE1}, representing it as an approximative limiting case when fluid flow and orientational dynamics decouple, was discussed in detail in~\cite{reinken_derivation_2018}. 
Note, however, that the dynamics defined by Eq.~\eqref{GSHE1} as it stands does not resolve the hydrodynamic flow explicitly but does contain hydrodynamic interactions via the transport coefficients and may therefore be considered a hybrid model in between the realms of dry and wet active matter. 
Its applicability is only given, strictly speaking, in the presence of a momentum sink such as a substrate. 
Its phenomenology is summarized in this section, cf.~Figs.~\ref{CompTurb},\ref{lengthScale}.

Eq.~\ref{GSHE1} has one fixed point~$\abs{\vec{P}} = 0$ corresponding to a disordered isotropic state and for $\alpha < 0$  a family of fixed points with $\abs{\vec{P}} > 0$ describing dynamic states with local polar order. 
A linear stability analysis shows that an unstable band of modes appears for $\Gamma_2<0$ when $4 \alpha <|\Gamma_2|^2/|\Gamma_4|$ where the isotropic fixed point becomes unstable and is replaced by a square vortex lattice for small values of~$\lambda_0$~\cite{dunkel_minimal_2013}. 
The periodicity of the lattice, i.e.~the vortex size, is given by the fastest growing mode corresponding to a wavelength $\Lambda \simeq \sqrt{\Gamma_4/\Gamma_2}$ obtained in the linear stability analysis. 
The stability of the square lattice was assessed by amplitude equations in~\cite{james_turbulence_2018}. 
In the presence of strong enough nonlinear advection, i.e.~if the parameter $\lambda_0$ is larger than a critical value, the regular periodic vortex lattice is destabilized and a new chaotic dynamic state with irregular vortex dynamics, called \textit{mesoscale turbulence}, emerges. 
Mesoscale turbulence is still characterized by a distinct length scale. 
In contrast to conventional turbulence in Newtonian fluids at high Reynolds numbers, energy spectra at large scales are non-universal, but depend on finite size effects and physical parameters~\cite{bratanov_new_2015}. 
By increasing the advection strength~$\lambda_0$ even further, a new hexagonal vortex pattern arises which spontaneously breaks the clockwise and anticlockwise rotation symmetry of the vortices. 
In distinction with classical pattern formation, the dominant length scale describing the hexagonal vortex lattice is given by the neutral mode of the dissipation from the linear stability analysis~\cite{james_turbulence_2018}.

The mesoscale turbulence state~--~irregular collective motion of creation and annihilation of vortices of a similar size~--~was found in suspensions of motile bacteria~\cite{wensink_emergent_2012}. 
A snapshot gallery~(2D slices of a 3D system) of measured fluid flow (moving fluorescent tracer, via particle tracking velocimetry~[PTV]), bacterial velocity (PIV) and numerical simulations indicates a good agreement~[Fig.\ref{CompTurb}(a-d)].
In particular, the extensive comparison of the dynamics of dense \emph{B.~subtilis} suspensions with numerical simulations of Eq.~\ref{GSHE1} in two and three spatial dimensions revealed that quantitative agreement at the level of velocity correlations, auto-correlation functions and energy spectra by simply adjusting the typical vortex size~$\Lambda$, the advection strength~$\lambda_0$ and the coefficient of the linear term~$\alpha$~\cite{wensink_emergent_2012,dunkel_fluid_2013}, cf.~Fig.\ref{lengthScale}(a-b).

There are different particle-based approaches addressing the dynamics of wet self-propelled rods.
The interplay of shape and hydrodynamics was studied in~\cite{theers_clustering_2018}.
An effective active particle model~--~a particle-based analogue to Eq.~\ref{GSHE1}~--~based on the competition of short range polar alignment and anti-alignment at larger distances that mimic the hydrodynamic interactions had been proposed: particle based simulations and systematic coarse-graining of the equations of motion revealed a variety of patterns from square vortex lattices and mesoscale turbulence to hexagonal vorticity arrays~\cite{grossmann_vortex_2014,grossmann_pattern_2015} [Fig.~\ref{CompTurb}(e)]. 

The typical vortex size is related by the fastest growing mode~$\Lambda$ is selected by the interplay between long-range hydrodynamics, short range alignment and self-propulsion. 
The properties of this new type of turbulence depend on microscopic details of the swimmer and their microscopic interactions. 
It is therefore a property of the active fluid itself. 
For increasing self-propulsion speed ($v_0$ activity) the vortex size increases and saturates at large $v_0$ as predicted by the theory~\cite{heidenreich_hydrodynamic_2016} and confirmed experimentally~\cite{sokolov_physical_2012}. 
In the limit of large self-propulsion speeds the vortex size is proportional to the length of individual rods. 
Such a dependence was recently found in experiments with \emph{B.~subtilis}~\cite{ilkanaiv_effect_2017}.
Recently, the interaction of~(mesoscale) turbulent \emph{B.~subtilis} suspension with an array of pillars was investigated experimentally:~a square lattice of vortices can be stabilized if the pillar-spacing roughly matches the characteristic length-scale indicated by the minimum of the correlation functions~\cite{nishiguchi_engineering_2018}.
This finding, in turn, substantiate the existence of an intrinsic length scale in mesoscale turbulence.

In summary, we note that long-range hydrodynamic interactions among wet self-propelled rods, reviewed in this section, lead to several novel phenomena in addition to the emergent collective behavior of dry self-propelled rods as reviewed in Section~\ref{sec:drySPR}, in particular mesoscale turbulence and vortex lattices appear as
possible large-scale patterns.

%%%%%%%%%%%%%%%%%%%%%%%%%%%%%%%%%%%%%%%%%%%%%%%%%%%%%%%
%%%%%%%%%%%%%%%%%%%%%%%%%%%%%%%%%%%%%%%%%%%%%%%%%%%%%%%

\section{SUMMARY, OUTLOOK AND FUTURE CHALLENGES}

We have reviewed experiments, models, theory and simulation results for self-propelled rods as a paradigmatic class of active matter.
The main parts of this review dealt with (i) the dynamics of ensembles of dry self-propelled rods, i.e.~self-propelled rods moving on a surface and having mostly short-range, contact-related steric or alignment interactions among each other and (ii) the collective motion of wet-self propelled rods, i.e.~rods swimming in a fluid that also are coupled through long-range dynamic interactions mediated by the flow field in the solvent.  
We close with a list of main results as well as future topics in the field of self-propelled rods.

Altogether, the reviewed research on self-propelled rods has already brought to light many surprising phenomena, a rich variety of patterns and dynamical states related to the nonequilibrium nature of this form of active matter. 
The achieved consensus is a good starting point for a deeper exploration of the fascinating world of biological self-organization. 
Future research will open the door to many important applications in material science and medicine.

\vspace{0.2cm}
\fboxsep0pt
\noindent
\colorbox{hellgrau}{
\begin{minipage}{0.988\textwidth}
\vspace{0.4cm}

\begin{center}
		
\parbox[c]{0.9\textwidth}{

	\vspace{0.16cm}

	\normalsize{\bfseries{\sffamily{\blue{\bfseries{\hspace*{-0.02cm}SUMMARY POINTS}}}}}

	\normalsize
	\normalfont

	\vspace{0cm}

 \begin{enumerate}
 \item Dry self propelled rods dominated by steric interaction can organize into polar clusters at low densities; the characteristic cluster size distributions are well reproduced by kinetic fragmentation-coagulation equations. 
 \item  Dry self-propelled rods with steric interaction at intermediate and high density show laning and formation of large clusters and polar bands.  
 \item  Dry self-propelled rods in quasi two dimensions, i.e.~rods that can slide over each other, are dominated by nematic alignment interactions and thus form nematic bands that are well described by coarse-grained continuum field equations. 
 \item  Wet self-propelled rods at high densities exhibit predominantly a state
 of mesoscale turbulence, i.e.~a disordered lattice of vortices with chaotic dynamics and a typical selected length-scale (average distance between vortices) that is well described by a continuum model. It can be derived from particle-based descriptions with short-ranged polar interaction and long-ranged interaction mediated through the flow field generated by the moving rods.
 \item Density, activity and shape are generic parameters which are important for all self-propelled systems. The knowledge of the specific nature of interactions between rods is essential for both the dry and the wet case as they often determine the emerging patterns or collective motion type~(polar or nematic). The relevant macroscopic order parameters and the symmetry of the coarse-grained equation will crucially depend on emergent symmetries. 
 \end{enumerate}

}

\end{center}

\vspace{-0.2cm}
\end{minipage}
}
\vspace{0.2cm}

\vspace{0.2cm}
\fboxsep0pt
\noindent
\colorbox{hellgrau}{
\begin{minipage}{0.988\textwidth}
\vspace{0.4cm}

\begin{center}
		
\parbox[c]{0.9\textwidth}{

	\vspace{0.16cm}

	\normalsize{\bfseries{\sffamily{\blue{\bfseries{\hspace*{-0.02cm}FUTURE ISSUES}}}}}

	\normalsize
	\normalfont

	\vspace{0cm}
	
\begin{enumerate}
\item Little is known about self-propelled rods in three dimensions~--~the transition from quasi-two dimensional systems representing monolayers or thin stripes of rods to multilayer systems is an important aspect of bacterial aggregation and biofilm formation yet to be explored.  
\item In many living systems additional aspects like biochemical communication, chemotaxis, switching between passive and (often multiple) active states, trail following, gene expression, synchronization of internal clocks or adhesion between active rods will have important implications for the dynamics and functionality of collective motion of self-propelled rods and will inspire improved and extended models as well as careful experimental studies. 
\item While first successful attempts to control the often chaotic and unpredictable dynamics of rods by confinement, added heterogeneities or external fields have been presented, a full understanding and exploitation of the potential of active rod assemblies as novel materials will require a better characterization of the transitions between different patterns together with a coordinated experimental effort.  
\item Descriptions related to self-propelled rods may be used for a better understanding of medical applications regarding e.g.~the dynamics of tissues and the spreading of cancerous cells. 
\item Tools from the emerging fields of data-driven modelling and machine learning are likely to be used routinely for better classifications of the multitude of possible dynamical states as well as for direct determination of the rules of motion and interaction from sufficiently large amounts of measurements. 
\end{enumerate}

}

\end{center}

\vspace{-0.2cm}
\end{minipage}
}
\vspace{0.2cm}

%%%%%%%%%%%%%%%%%%%%%%%%%%%%%%%%%%%%%%%%%%%%%%%%%%%%%%%%%%%%%%%%%%%%%%%%%%%%%%%%%%%%%%%%%%%%%%%%%%%%%%%%%%5
%
% appendix
%
%%%%%%%%%%%%%%%%%%%%%%%%%%%%%%%%%%%%%%%%%%%%%%%%%%%%%%%%%%%%%%%%%%%%%%%%%%%%%%%%%%%%%%%%%%%%%%%%%%%%%%%%%%5

% % Disclosure
% % If the authors have noting to disclose, the following statement will be used: 
% \section*{DISCLOSURE STATEMENT}
% The authors are not aware of any affiliations, memberships, funding, or financial holdings that might be perceived as affecting the objectivity of this review. 

% Acknowledgements
\section*{ACKNOWLEDGMENTS}

MB and SH thank DFG for support through SFB910 and the DFG middle-east program~(project no. 396653815) for financial support of this research. 
FP acknowledges financial support from Agence Nationale de la Recherche via project \textit{BactPhys}, Grant ANR-15-CE30-0002-01. 
We benefited from discussions and collaborations with Gil Ariel, Igor Aranson, Avraham Be{\textquotesingle}er, Hugues Chat\'{e}, Andreas Deutsch, Francesco Ginelli, Sabine Klapp, Henning Reinken, Holger Stark, J\"orn Dunkel, Pawel Romanczuk and Lutz Schimansky-Geier.

%%% 
% bibliography
%%% 

\end{document}